\documentclass[10pt,a4paper]{article}
\usepackage{jheppub}

\usepackage[T1]{fontenc}
\usepackage[utf8]{inputenc}
\usepackage{amsmath}
\usepackage{amsfonts}
\usepackage{amssymb}
\usepackage[english]{babel}
\usepackage[dvipsnames]{xcolor}
\definecolor{linkcolor}{rgb}{.17578125,.1875,.5703125}
\usepackage{hyperref}
\hypersetup{
    colorlinks=true,
    linkcolor=linkcolor,
    filecolor=magenta,      
    urlcolor=linkcolor,
    citecolor=BrickRed
}
\usepackage{graphicx}
\usepackage{subcaption}
\usepackage{float}
\usepackage{bbm}
\usepackage{comment}
\usepackage{marvosym}
\usepackage{pifont}
\usepackage{slashed}
\usepackage{cancel}
\usepackage{braket}
\usepackage{parskip}

\usepackage{cleveref}

\newcommand{\nn}{\nonumber}

\newcommand{\dd}{\mathrm{d}}

\newcommand{\pvec}{\mathbf{p}}
\newcommand{\vlvec}{\mathbf{v}_{\ell}}

\newcommand{\kvec}{\mathbf{k}}

\newcommand{\nvec}{\mathbf{n}}

\newcommand{\qedl}{\text{QED$_\text{L}$}}

\newcommand{\qedr}{\text{QED$_\text{r}$}}

\newcommand{\qedc}{\mathrm{QED}_{\mathrm{C}}}
\newcommand{\qedm}{\mathrm{QED}_{\mathrm{M}}}

\crefformat{section}{section~#2#1#3}
\crefmultiformat{section}{sections~#2#1#3}
    { and~#2#1#3}{, #2#1#3}{ and~#2#1#3}
\crefrangeformat{section}{sections~#3#1#4\,--\,#5#2#6}
\crefformat{subsection}{section~#2#1#3}
\crefmultiformat{subsection}{sections~#2#1#3}
    { and~#2#1#3}{, #2#1#3}{ and~#2#1#3}
\crefrangeformat{subsection}{sections~#3#1#4\,--\,#5#2#6}
\crefformat{subsubsection}{section~#2#1#3}
\crefmultiformat{subsubsection}{sections~#2#1#3}
    { and~#2#1#3}{, #2#1#3}{ and~#2#1#3}
\crefrangeformat{subsubsection}{sections~#3#1#4\,--\,#5#2#6}
\crefformat{figure}{figure~#2#1#3}
\crefmultiformat{figure}{figures~#2#1#3}
    { and~#2#1#3}{, #2#1#3}{ and~#2#1#3}
\crefrangeformat{figure}{figures~#3#1#4\,--\,#5#2#6}
\crefformat{table}{table~#2#1#3}
\crefmultiformat{table}{tables~#2#1#3}
    { and~#2#1#3}{, #2#1#3}{ and~#2#1#3}
\crefrangeformat{table}{tables~#3#1#4\,--\,#5#2#6}
\crefformat{equation}{eq.~(#2#1#3)}
\crefmultiformat{equation}{eqs.~(#2#1#3)}{ and~(#2#1#3)}{, (#2#1#3)}{ and~(#2#1#3)}
\crefrangeformat{equation}{eqs.~(#3#1#4)--(#5#2#6)}

\newcommand{\cern}{Theoretical Physics Department, CERN, 1211 Geneva 23, Switzerland}
\newcommand{\edin}{School of Physics and Astronomy, University of Edinburgh, Edinburgh EH9 3FD, United Kingdom}
\newcommand{\lund}{Division of Particle and Nuclear Physics, Department of Physics, Lund University, Box 43, SE 221 00 Lund, Sweden}

\title{QED$_\text{r}$: a finite-volume QED action with redistributed spatial zero-momentum modes}

\author[a,b]{Matteo Di Carlo,}
\author[b]{Maxwell T. Hansen,}
\author[b,c]{Nils Hermansson-Truedsson}
\author[b]{and Antonin Portelli}

\affiliation[a]{\cern}
\affiliation[b]{\edin}
\affiliation[c]{\lund}

\emailAdd{matteo.dicarlo@cern.ch, maxwell.hansen@ed.ac.uk,  nils.hermansson-truedsson@ed.ac.uk, antonin.portelli@ed.ac.uk}

\abstract{
We present a finite-volume QED action designed to improve the infinite-volume extrapolation of hadronic observables in precision lattice QCD{+}QED calculations. The action proposed in this work, which we call $\qedr$, can be seen as a particular case of the infrared-improved QED actions introduced by Davoudi et al.~in 2019, and is specifically designed to remove kinematics-independent finite-volume corrections that appear at $\mathrm{O}(1/L^3)$ in the commonly used $\qedl$ formulation, where $L$ is the spatial extent of the physical volume. For a number of key observables, these effects depend on the internal structure of the hadrons and are difficult to evaluate non-perturbatively, making an analytical subtraction of the finite-volume effects impractical. We explicitly study the $\qedr$ electromagnetic finite-size effects on hadron masses and leptonic decay rates, relevant for Standard Model precision tests using the Cabibbo-Kobayashi-Maskawa matrix elements. In addition, we propose methods to remove the kinematics-dependent $\mathrm{O}(1/L^3)$ effects in leptonic decays. The removal of such contributions, shifting the leading contamination to $\mathrm{O}(1/L^4)$, will help to reduce the systematic uncertainties associated with finite-volume effects in future lattice QCD{+}QED calculations.

}

\preprint{CERN-TH-2025-001}

\begin{document}

\maketitle

\hfill

%sections

\section{Introduction}

Scenarios for new physics beyond the Standard Model (SM) can be constrained by precision tests that combine theoretical predictions with experimental measurements. For observables where the non-perturbative dynamics of the strong interactions is important, lattice quantum chromodynamics (QCD) plays a central role in providing reliable SM theory estimates,\footnote{See, for example, the latest summary of the Flavour Lattice Averaging Group~\cite{FlavourLatticeAveragingGroupFLAG:2024oxs}.} and over the last two decades the relative precision of lattice QCD calculations in the isospin-symmetric limit has reached the sub-percent level for many observables. This has necessitated the inclusion of isospin-breaking effects due to the up and down quark mass difference and electromagnetism, which are expected to contribute at the percent level. In these cases, quantum electrodynamics (QED) must be included in the lattice calculations.

Examples of observables for which isospin-breaking corrections are relevant and are investigated via lattice calculations include the hadronic mass spectrum~\cite{deDivitiis:2013xla,BMW:2014pzb,Clark:2022wjy,Frezzotti:2022dwn,RCstar:2022yjz,Segner:2023igh}, the muon anomalous magnetic moment~\cite{Giusti:2017jof,RBC:2018dos,Giusti:2019xct,Aoyama:2020ynm,Borsanyi:2020mff,Altherr:2022fqa,Biloshytskyi:2022ets,Chao:2023lxw,Boccaletti:2024guq,Djukanovic:2024cmq,Parrino:2025afq} and leptonic decay rates of pseudoscalar mesons~\cite{Carrasco:2015xwa,Giusti:2017dwk,DiCarlo:2019thl,Boyle:2022lsi,Christ:2023lcc}. The latter can be used to determine Cabibbo-Kobayashi-Maskawa (CKM) matrix elements, which play a central role in flavour-physics precision tests.

Lattice QCD calculations are generically performed on a discretized spacetime of finite extent, with periodic boundary conditions imposed on the quark and gluon fields. For concreteness, one typically takes three spatial extents of equal length $L$ and a longer temporal extent of length $T$. When QED is included directly in such calculations, a set of periodic boundary conditions must also be applied to the photon fields. However, the long-range nature of QED makes the definition of the photon action and hence of charged states in a finite volume with periodic boundary conditions highly non-trivial~\cite{Kronfeld:1990qu,Wiese:1991ku,Kronfeld:1992ae,Polley:1993bn,Duncan:1996be,Duncan:1996xy,Hayakawa:2008an,Endres:2015gda,Lucini:2015hfa,Davoudi:2018qpl}. The issue can be understood conceptually from Gauss' law, which implies that the flux through a Gaussian surface must be non-zero if that surface encloses non-zero charge. This contradicts the fact that naive periodic boundary conditions yield zero flux through the boundary of the volume. The problem is related to the zero-momentum modes of the massless physical photons, and also to large gauge transformations in the QED action~\cite{Hayakawa:2008an,Davoudi:2018qpl}.

Several methods to circumvent this issue have been published to date. In the QED$_{\textrm{M}}$ prescription~\cite{Endres:2015gda,Bussone:2017xkb}, a mass $m_\gamma$ is given to the photon field. The ordered double limit of $L \to \infty$ followed by $m_\gamma \to 0$ is then taken in order to obtain physical predictions. In $\qedc$~\cite{Kronfeld:1990qu,Wiese:1991ku,Kronfeld:1992ae,Polley:1993bn,Lucini:2015hfa}, an alternative periodicity involving charge conjugation is imposed. This removes the zero-momentum mode of the photon while preserving gauge invariance and locality, but breaks charge and flavour conservation at finite $L$. In $\qedl$~\cite{Hayakawa:2008an}, the spatial zero-momentum modes of the photon are removed on each time-slice, thus rendering the theory non-local. Despite this non-locality, $\qedl$ remains the method most commonly used in practical lattice calculations~\cite{BMW:2014pzb,Davoudi:2018qpl,DiCarlo:2019thl,Boyle:2022lsi,Boccaletti:2024guq}.

A complementary approach to the finite-volume formulations of QED, sometimes referred to as QED${}_\infty$, involves expressing QED corrections with infinite-volume kernel functions convoluted with hadronic matrix elements~\cite{Asmussen:2016lse,Feng:2018qpx,Biloshytskyi:2022ets,Christ:2023lcc}. Finite-volume lattice calculations are used to determine the latter at small to intermediate length scales. For large separations the infinite-volume hadronic matrix elements can be modeled using the dominance of single-hadron states in a spectral decomposition.
To reach a final result, the prescription for separating long- and short-distance effects must be varied and studied in detail on a case-by-case basis~\cite{Feng:2018qpx,Christ:2023lcc}.

All QED prescriptions outlined above are expected to recover the same results for any observable in the infinite-volume limit. However, the finite-volume dependence of a given observable can vary significantly across prescriptions. For $\qedc$ and $\qedl$, the volume dependence scales as powers of $1/L$, along with potential logarithmic infrared divergences. In $\qedm$ the effects are exponentially suppressed for $m_\gamma \gg 1/L$, but polynomial dependence arises when this hierarchy is violated~\cite{Clark:2022wjy}. In QED${}_\infty$, the scaling is exponentially suppressed, with the scale of suppression depending delicately on the separation of distances scales~\cite{Asmussen:2016lse,Feng:2018qpx,Biloshytskyi:2022ets,Christ:2023lcc}.

To better understand this landscape of possible schemes, the authors of ref.~\cite{Davoudi:2018qpl} proposed a generalization of the $\qedl$ action to improve the infrared behaviour of hadronic observables while retaining the technical simplicity of implementing the scheme. The key idea is to modify multiple low-lying momentum modes of the photon action, to achieve certain desired behavior in the $L$ dependence of hadronic observables without affecting the $L \to \infty$ limit. As we discuss below, the $\qedr$ formulation introduced in this work is a specific subset of the family of choices outlined in ref.~\cite{Davoudi:2018qpl}. The key distinction here is that we have identified the $1/L^3$ contamination as particularly problematic, in short because it depends on the internal structure of the hadron and is difficult to evaluate non-perturbatively, and because it is known to be numerically enhanced for certain observables. The $\qedr$ prescription sets this term to zero.

Before turning to the main text, we close this introduction by summarising the motivation for the $\qedr$ prescription in more detail, beginning with the ratio of light- and strange-quark leptonic decays, and the corresponding ratio of CKM matrix elements.

Within the SM, the ratio $|V_{us}|/|V_{ud}|$ can be obtained from the experimentally determined ratio of kaon and pion leptonic decay rates, together with theory predictions for the isospin-symmetric decay constants, and the isospin-breaking effects contained in the dimensionless parameter $\delta R_{K\pi} = \delta R_K-\delta R_\pi$. Here $\delta R_P$ is the isospin-breaking effect to the inclusive decay rate, $\Gamma [P\rightarrow \ell \nu _\ell (\gamma)] = \Gamma ^{\textrm{tree}}_P (1+\delta R_P)$ with $\Gamma ^{\textrm{tree}}_P$ the isospin-symmetric result.
In refs.~\cite{Giusti:2017dwk,DiCarlo:2019thl,Boyle:2022lsi}, $\delta R_{K\pi}$ was calculated in lattice QCD+$\qedl$ from matrix elements describing leptonic decays with muons in the final state, $K\rightarrow \mu\nu_\mu$ and $\pi \rightarrow \mu \nu _\mu$. Only the virtual QED corrections to the inclusive decay rates were calculated on the lattice, with the real radiative part calculated perturbatively with a photon mass regulator as outlined in ref.~\cite{Carrasco:2015xwa}. The validity of computing the real radiative part perturbatively in the case of pion and kaon decays into muons was later confirmed through dedicated numerical lattice simulations~\cite{Desiderio:2020oej,Frezzotti:2020bfa}. To remove finite-volume artefacts from the calculation in ref.~\cite{Boyle:2022lsi}, the leading volume dependence was determined analytically and subtracted. Specifically, terms scaling as $1/L$ and $1/L^2$ were removed while the neglected $1/L^3$ scaling was incorporated into the systematic uncertainty of the final result~\cite{Lubicz:2016xro,DiCarlo:2021apt}. 
The quantity $\delta R_{K\pi}$ was found to be~\cite{Boyle:2022lsi}
\begin{align}\label{eq:deltaRKPi}
\delta R_{K\pi} = -0.0086 (13)_{\textrm{stat.+sys.}} (39)_{\textrm{vol.}}
\, ,
\end{align}
where the first error contains statistical uncertainties as well as systematics except for those associated to the volume dependence. A full separation of the error budget can be found in ref.~\cite{Boyle:2022lsi}. It should be noted that the uncertainty associated to the volume amounts to roughly $50\%$ of the central value of $\delta R_{K\pi}$. Propagating this to $|V_{us}/V_{ud}|$ prohibits precision tests of CKM unitarity using the first-principle result for $\delta R_{K\pi}$ from ref.~\cite{Boyle:2022lsi}. 

The large error arises due to the truncation of the large $L$ expansion of the finite-volume virtual decay rate $\Gamma _0 (L)$. This is conveniently encoded in a function called $Y^{\textrm{L}}(L)$, defined via
\begin{align}
\Gamma_0 (L)= \Gamma ^{\textrm{tree}}_P \left [ 1+2\, \frac{\alpha}{4\pi }\, Y^{\textrm{L}}(L) \right ] \,.
\end{align}
The expansion can then be written as
\begin{align}
Y^{\textrm{L}}(L) = \tilde{Y}^{\textrm{L}} (L) + Y^{\textrm{L}}_{0} + \frac{Y^{\textrm{L}}_{1}}{L}
+ \frac{Y^{\textrm{L}}_{2}}{L^2} + \frac{Y^{\textrm{L}}_{3}}{L^3} + O \left[ \frac{1}{L^4},e^{-m_\pi L}\right] \, ,
\end{align}
and the coefficients entering this expansion are discussed in detail in refs.~\cite{Lubicz:2016xro,DiCarlo:2021apt}. 

The essential points are that $\tilde{Y}^{\textrm{L}} (L)$, $Y^{\textrm{L}}_{0}$, $Y^{\textrm{L}}_{1}$ and $Y^{\textrm{L}}_{2}$ are fully known analytically, while $Y^{\textrm{L}}_{3}$ is only known in the point-like limit, in which the structure of the pseudoscalar is ignored. A full determination of $Y^{\textrm{L}}_{3}$ is highly non-trivial, in particular as it depends on integrals over branch cuts in various related scattering amplitudes. In ref.~\cite{Boyle:2022lsi} the point-like value of $Y^{\textrm{L}}_{3}$ was used as a systematic error on the single-volume calculation of $\delta R_{K\pi}$. The value is significantly enhanced, and this led to the large uncertainty in \cref{eq:deltaRKPi} quoted above. The $\qedr$ prescription is thus designed to specifically remove the nontrivial structure dependence in $Y^{\textrm{L}}_{3}$. We emphasize that, regardless of the QED prescription adopted, a numerical extrapolation over a range of volumes remains essential, as was done in the case of leptonic decays in refs.~\cite{Giusti:2017dwk,DiCarlo:2019thl}. Analytical understanding of finite-volume scaling can then guide the choice of fit function~\cite{Boyle:2022lsi}.

Irrespective of this motivation we expect $\qedr$ will be useful for other observables, for example for hadron masses as well as the hadronic vacuum polarisation contribution to the muon anomalous magnetic moment~\cite{Bijnens:2019ejw,Hermansson-Truedsson:2024mey}. Taking the former as another example, we define $\Delta m_{P}^2 (L)$ as the difference between the finite- and infinite-volume squared pseudoscalar masses. One can then identify a similar expansion of the form~\cite{DiCarlo:2021apt}
\begin{align}
	\frac{\Delta m_{P}^2 (L)}{e^2\, m_P^2}
	& =
\frac{\delta m_1^\textrm{L}}{L} +
\frac{\delta m_2^\textrm{L}}{L^2} +
\frac{\delta m_3^\textrm{L}}{L^3}
+{O} \left[\frac{1}{(m_{P}L)^4},e^{-m_{P}L}\right]
	\, ,
\end{align}
where the coefficients $\delta m^\textrm{L}_i$ can be found in ref.~\cite{DiCarlo:2021apt}. As was the case with $Y^{\textrm{L}}_{3}$, the coefficient
$\delta m_3^\textrm{L}$ contains
non-trivial structure dependence defined by an integral over the branch cut in the Compton scattering amplitude $P+\gamma \rightarrow P+\gamma$.\footnote{Here we primarily have in mind derivations in which quantities are represented using a skeleton expansion with irreducible vertex functions. This allows one to express finite-volume effects in terms of charges, form factors, and other matrix elements. An alternative approach is to use effective field theory to directly calculate the coefficients in terms of low-energy constants. See e.g. refs.~\cite{Colangelo:2005gd,Colangelo:2006mp,Bijnens:2014dea,Bijnens:2017esv} for EFT-specific finite-volume effects in QCD and refs.~\cite{Hayakawa:2008an,BMW:2014pzb,Davoudi:2014qua,Lubicz:2016xro,Bijnens:2019ejw} for EFT-specific isospin-breaking effects.
}

Common to the two expansions above is the appearance of geometric coefficients $c_j$, defined as
\begin{align}
c_j = \left( \left. \sum _{\mathbf{n}\neq \mathbf{0}}\right. -\int d^3 \mathbf{n}\right) \frac{1}{|\mathbf{n}|^j} \, .
\end{align}
In particular the coefficient $c_0 = -1$ appears in both $\delta m^\textrm{L}_3$ and $Y^{\textrm{L}}_{3}$ and is non-zero due to the removed zero-mode $\mathbf{n}=\mathbf{0}$ in the sum above. This extra level of detail lets us articulate the spirit of the $\qedr$ prescription more precisely. The motivation (and the defining property) of the scheme is simply to modify the sum such that $c_0$ is set to zero.

The remainder of this manuscript is structured as follows. The $\qedr$ prescription is defined in~\cref{sec:genqedr}.
The large-volume expansion for the hadron masses is presented in~\cref{sec:selfenergy}. In~\cref{sec:lepdec} we proceed to virtual QED corrections to leptonic decay rates, and finally provide conclusions and an outlook in~\cref{sec:conclusions}. In~\cref{app:qedcfvcoeffs} a small study of the numerical values of finite-volume coefficients is presented. In all sections, unless otherwise stated, we consider a continuous Euclidean spacetime of infinite time extent and the spatial dimensions constrained to have length $L$ with periodic boundary conditions.
\section{Definition of $\qedr$}
\label{sec:genqedr}

As discussed in the introduction, the purpose of this work is to introduce and explore the consequences of a new definition of finite-volume QED that we call $\qedr$, corresponding to a specific choice within the class of actions in ref.~\cite{Davoudi:2018qpl}. The spirit of the definition is that, instead of simply removing the spatial zero-momentum mode as in $\qedl$, we instead advocate \emph{redistributing} the mode to neighbouring non-zero momenta. For concreteness we mainly restrict attention to the approach of modifying a set of modes with fixed magnitude, i.e. a set of $\kvec$ satisfying $\kvec^2 = (2 \pi/L)^2 R^2$ for some positive integer $R^2$.

At the level of the finite-volume Euclidean photon propagator, the definition reads
\begin{equation}
\label{eq:qedr_propagator}
D_{\mu\nu}^\mathrm{r}(x) = \frac{1}{L^3}\sum_{\kvec\neq\mathbf{0}}\int\frac{\dd k_0}{2\pi} \,e^{ikx} \, \frac{\delta_{\mu\nu}}{k_0^2+\kvec^2} \big[1+h(\nvec,R)\big] \bigg \vert_{\textbf n = \kvec L/(2 \pi)}\,.
\end{equation}
This can be viewed as the $\qedl$ propagator:
\begin{equation}
D_{\mu\nu}^\mathrm{L}(x) = \frac{1}{L^3}\sum_{\kvec\neq\mathbf{0}}\int\frac{\dd k_0}{2\pi} \,e^{ikx} \, \frac{\delta_{\mu\nu}}{k_0^2+\kvec^2}\,,
\end{equation}
plus an additional term containing $h(\nvec,R)$. We emphasize that both propagators are defined in Feynman gauge and, in both cases, fixing to this particular gauge before modifying the spatial momentum modes is  part of the definition of the scheme.

The function $h(\nvec,R)$ only takes on non-zero values when $\nvec^2 = R^2$ and can be written as
\begin{equation}
h(\nvec,R) = w(\nvec,R) \, \delta_{\nvec^2,R^2}\,,
\end{equation}
where $w(\nvec,R)$ is a weight function that is only defined for $\nvec$ satisfying $\vert \nvec \vert = R$.
It is convenient to define $\mathcal{S}_R$ as the set of all such integer three-vectors:
\begin{equation}
\mathcal{S}_R = \big\{\nvec \in \mathbb{Z}^3 \ \big| \ |\nvec|= R \big\}\,.
\end{equation}
Then $w(\nvec,R)$ is defined for $\nvec \in \mathcal{S}_R$ and is required to satisfy the normalization condition
\begin{equation}
\sum_{\nvec\in\mathcal{S}_R} w(\nvec,R)=1\,,
\end{equation}
implying
\begin{equation}
\sum_{\kvec \neq \mathbf{0}} h(\nvec,R) \bigg \vert_{\textbf n = \kvec L/(2 \pi)} = 1\,.
\end{equation}
This property ensures the removal of all $1/L^3$ terms that are proportional to $c_0$ in $\qedl$, and this is the defining property of $\qedr$ as a subset of the family of improved actions proposed in ref.~\cite{Davoudi:2018qpl}.

We note that, in principle, the definition of the $\qedr$ propagator in~\cref{eq:qedr_propagator} can be extended to the case of zero mode redistribution over $N$ shells with radii $\boldsymbol{R}=\{R_1,R_2,\dots, R_N\}$ by replacing the weight function $h(\nvec,R)$ with
\begin{equation}
h(\nvec,\boldsymbol{R}) = \sum_{\alpha\in\boldsymbol{R}}\omega(\alpha) h(\nvec,\alpha)\,,
\end{equation}
where $\omega(\alpha)$ denotes the weight assigned to the shell $|\nvec|=\alpha$ defined such that $\sum_\alpha\omega(\alpha)=1$. We will not pursue this multiple-shell option further in this work.

The propagator in~\cref{eq:qedr_propagator} is derived from the Feynman-gauge Euclidean action
\begin{equation}
\label{eq:qedr_action_feynman}
S_\mathrm{r}[\hat{A}_\mu] = \frac{1}{2L^3}\sum_{\kvec\neq\boldsymbol{0}}\int\frac{\dd k_0}{2\pi} \, \hat{A}_\mu(k)^* \, \bigg[\frac{ \delta^{\mu\nu} k^2 }{1+h(\nvec,R)}\bigg]\, \hat{A}_\nu(k) \,,
\end{equation}
with $\hat{A}_\mu(k)$ defined as the Fourier-transformed photon field and ${\hat{A}_\mu(k)^*=\hat{A}_\mu(-k)}$. As noted in ref.~\cite{Davoudi:2018qpl}, which provides a generalized form of~\cref{eq:qedr_action_feynman} applicable to any gauge, the positivity of the action requires the weight function $h(\nvec, R)$ to satisfy ${h(\nvec, R) > -1}$. Combining this with the normalization condition on the weights, we infer that ${|w(\nvec, R)| < 1}$.

In the case of an isotropic observable, in which no particular directions e.g.~in momentum space are singled out, it is most natural to assign equal weights to the modes $\nvec\in\mathcal{S}_R$ with values $w(\nvec,R) = 1/r_3(R^2)$, where the function $r_3(R^2) = \sum_{\nvec}\delta_{\nvec^2,R^2}$ counts the number of integer three-vectors with magnitude $R$.  In the notation of ref.~\cite{Davoudi:2018qpl} (Sec.~IV.A), this corresponds to $w_{|\nvec|^2}=\delta_{\nvec^2,R^2}/r_3(R^2)$. In the following, we will study the simplest implementation of $\qedr$, which we call \emph{minimal $\qedr$}, in which the isotropic redistribution of the zero mode is applied to the $R=1$ shell. This corresponds to assigning $w(\nvec,1)=1/6$ to the six modes in $\mathcal{S}_1$. Below we also use the shorthand $h(\nvec)=h(\nvec,1)$.

Applying the minimal $\qedr$ set-up to the propagator in~\cref{eq:qedr_propagator}, we deduce that finite-volume effects will arise from sum-integral differences of the form
\begin{equation}
\label{eq:Delta_k_qedr}
\Delta_\kvec^\mathrm{r} \, f(\kvec) = \bigg[\frac{1}{L^3}\sum_{\kvec\neq\boldsymbol{0}} - \int\frac{\dd^3\kvec}{(2\pi)^3}\bigg] \big(1+h(\nvec)\big) \bigg \vert_{\nvec = \kvec L/(2 \pi)}
f(\kvec)\,,
\end{equation}
where we defined the operator $\Delta_\kvec^\mathrm{r}$ acting on a generic function $f(\kvec)$. It is further convenient to define a dimensionless analog in which $\kvec = 2\pi \nvec/L$ is replaced by $\nvec$:
\begin{equation}
\label{eq:Delta_n_qedr}
\Delta_\nvec^\mathrm{r} F(\nvec) = \bigg[\sum_{\nvec\neq\boldsymbol{0}} - \int\dd^3\nvec\bigg]\big(1+h(\nvec)\big) F(\nvec)\,.
\end{equation}
In terms of this operator, we will see below that the $\qedr$ finite-volume effects relevant for this paper are parametrized by two new sets of geometric coefficients
\begin{equation}
\bar{c}_j = \Delta_\nvec^\mathrm{r} \frac{1}{|\nvec|^j}\,,\qquad
\bar{c}_j(\mathbf{v}) = \Delta_\nvec^\mathrm{r} \, \frac{1}{|\nvec|^j}\frac{1}{(1-\mathbf{v}\cdot \hat{\nvec})}\,,
\end{equation}
where $\mathbf{v}$ is a generic dimensionless vector that, for the study of leptonic decays in~\cref{sec:lepdec}, will become the velocity of the outgoing lepton in the finite-volume frame. Our statement above, that terms proportional to $c_0$ in $\qedl$ are removed in $\qedr$, is equivalent to the statement that $\bar{c}_0 = 0$.

This completes our general discussion of $\qedr$. We now turn to a detailed study of the consquences of this definition for finite-volume effects in QED corrections to hadronic observables.
\section{Pseudoscalar meson masses}\label{sec:selfenergy}

In this section we study the finite-volume corrections to the mass $m_P$ of a pseudoscalar meson $P$. The calculation of such effects has been performed to various orders and approximations in refs.~\cite{Hayakawa:2008an,BMW:2014pzb,Davoudi:2014qua,Lubicz:2016xro,Davoudi:2018qpl,DiCarlo:2021apt} using the $\qedl$ regularization and in ref.~\cite{Lucini:2015hfa}  for $\qedc$. Here we follow the approach outlined in ref.~\cite{DiCarlo:2021apt} and compute the finite-volume effects in $\qedr$, showing that contributions of $\mathrm{O}(1/L^3)$ vanish in this regularization.\footnote{For an alternative derivation, see also ref.~\cite{DiCarlo:2024lue}.}

\begin{figure}[t!]
	\centering
	\includegraphics[width=0.3\textwidth]{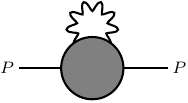}
	\caption{Diagrammatic representation of the mass shift in \cref{eq:massshiftcompton}, where the Compton tensor enters in the grey blob. The wiggly line corresponds to the photon.}
	\label{fig:sediagram}
\end{figure}

\begin{figure}[t!]
	\centering
	\includegraphics[width=0.5\textwidth]{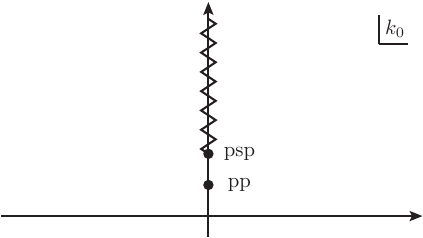}
	\caption{Analytical structure of the integrand function in~\cref{eq:massshiftcompton} in the upper half of the $k_0$-plane. The photon pole is here labelled as $\textrm{pp}$ and the pseudoscalar pole as $\textrm{psp}$.
    }
	\label{fig:comptonanalytic}
\end{figure}

Let us consider the meson $P$ in its rest frame. Denoting $p=(p_0,\boldsymbol{0})$ its four-momentum with $p^2=-m_P^2$ when evaluated on-shell, the finite-volume corrections can be obtained as~\cite{DiCarlo:2021apt}
\begin{equation}
\label{eq:massshiftcompton}
	\Delta m_P^2 (L)\big|_{\qedr} = -\frac{e^2}{2} \lim_{p^2 \to -m_P^2} \Delta^\mathrm{r}_{\mathbf{k}} \int \frac{dk_0}{2\pi} \, \frac{C_{\mu\mu}(p,k,-k)}{k^2}
    \, ,
\end{equation}
where $C_{\mu\nu}(p,k,-k)$ is the off-shell forward Compton scattering amplitude such that
\begin{equation}
    \lim_{p^2 \to -m_P^2} C_{\mu\mu}(p,k,-k) = T(k^2,k\cdot p) = \int \dd^4 x \, e^{-ikx} \, \langle P,\boldsymbol{0} | \mathrm{T} \big\{ J_\mu(x) J_\mu(0)\big\} | P,\boldsymbol{0} \rangle\,,
\end{equation}
and $\Delta_\mathbf{k}^\mathrm{r}$ denotes the sum-integral difference operator defined above in~\cref{eq:Delta_k_qedr} which includes the $\qedr$ reweighing of the spatial modes. This is depicted in~\cref{fig:sediagram}.
The decomposition of the Compton tensor $C_{\mu\nu}(p,k,-k)$ in terms of irreducible vertex functions by means of a skeleton expansion has been studied in ref.~\cite{DiCarlo:2021apt}, and would here correspond to a separation of the diagram in~\cref{fig:sediagram} into a sum of contributions.
The $k_0$ integral can be performed using Cauchy's theorem, deforming the integration contour in the upper half of the $k_0$ complex plane and encircling the singularities of the photon propagator and of the  Compton amplitude. The Compton amplitude has a pole at $k_0 = p_0 + i \omega_P(\kvec)$, where the energy function $\omega _{P}(\kvec) = \sqrt{\kvec ^2 + m_P^2}$, in correspondence to an on-shell intermediate pseudoscalar meson propagating between the two electromagnetic currents, as well as a multi-particle cut. The starting point of the cut depends on $\mathbf{k}$, which for $\mathbf{k}=\mathbf{0}$ is at $k_0 = 2im_\pi$. A visual representation of the analytical structure of the integrand of~\cref{eq:massshiftcompton} is given in~\cref{fig:comptonanalytic}, when $p_0 = im_P$.
%%%
%%%
The $k_0$ integration returns a function of the photon spatial momentum $\kvec$. Following ref.~\cite{DiCarlo:2021apt}, we can now substitute $\kvec=2\pi\mathbf{n}/L$ ($\mathbf{n}\in\mathbb{Z}^3$) and perform a large $L$ expansion, which yields
\begin{equation}
\label{eq:massrestframe}
	\Delta m_{P}^2 (L)\big|_\mathrm{\qedr} 
	 =
	e^2 m_P^2 \Bigg\{
	\frac{ \bar{c}_{2}}{4\pi^2\, m_P L}+\frac{\bar{c}_1 }{2\pi \, (m_P L) ^2}
	-\frac{\bar{c}_0}{(m_P L )^3}\bigg(\frac{\langle r^2_{P}\rangle  m_P^2}{3} \,+\, \mathcal{C}\bigg) 
+{O} \bigg[\frac{1}{(m_{P}L)^4}\bigg] 	\Bigg\} 
	\, ,
\end{equation}
 up to exponentially suppressed finite-volume corrections.
Here $\langle r_P^2\rangle$ is the squared charge radius of the meson, while $\mathcal{C}$ denotes the contribution of the integration over the multi-particle branch cut~\cite{DiCarlo:2021apt}
\begin{equation}
    \mathcal{C} = \frac{m_P}{2}\int_\textsf{cut} \frac{\dd k_0}{2\pi} \frac{T(k_0^2,i m_P k_0)}{k_0^2}\,.
\end{equation}
The finite-volume coefficients $\bar{c}_j$ appearing in the expression above are defined as
\begin{equation}
    \bar{c}_j = \Delta_\mathbf{n}^\mathrm{r} \, \frac{1}{|\mathbf{n}|^j}\,.
\end{equation}
The property of the $\qedr$ weights that $\sum_{\mathbf{n}}h(\mathbf{n})=1$ yields a general relation between the $\overline{c}_j$ in $\qedr$ and the $c_j$ in $\qedl$, namely
\begin{equation}\label{eq:qedlrrel}
    \overline{c}_j = c_j+1 \, . 
\end{equation}
In particular, this guarantees that the coefficient $\bar{c}_0$ appearing at $\mathrm{O}(1/L^3)$ vanishes,
\begin{equation}
    \bar{c}_0 = c_0 + 1 = 0 \, .
\end{equation}
In $\qedl$, the appearance of a non-zero coefficient $c_0$ is due to the removal of the spatial zero mode $\kvec=\boldsymbol{0}$. In $\qedr$ we can interpret the absence of such contribution at $\mathrm{O}(1/L^3)$ as a consequence of the redistribution of the zero mode over the neighbouring Fourier modes. When taking the infinite-volume limit, the Fourier space becomes denser as the volume grows and the weights of the neighbouring modes reproduce the contribution of the zero mode, which in $\qedl$ would be simply removed. It should be noted that the shifted coefficients in $\overline{c}_j$ in~\cref{eq:qedlrrel} affect all orders in the finite-volume expansion. Motivated by the need for a better understanding through order $1/L^3$ in practical data analyses, we do not consider higher-order effects in this paper. We leave a study of the tail of the expansion to future work.
\section{Leptonic decays}
\label{sec:lepdec}

\begin{figure}[t!]
	\centering
	\includegraphics[width=0.25\textwidth]{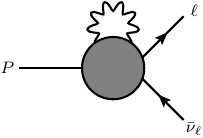}
	\caption{Diagrammatic representation of the leptonic decay, with the photon represented by the wiggly line. The grey blob contains the weak current mediating the decay.}
	\label{fig:pl2diagram}
\end{figure}

We consider now the leptonic decay $P^-\to \ell^-\bar{\nu}_\ell$ of a negatively charged pseudoscalar meson into a lepton and neutrino pair. In the rest frame of the decaying meson, energy and momentum conservation implies that the squared amplitude of the process $|\mathcal{M}|^2$ only depends on the spatial momentum of the final-state charged lepton $\mathbf{p}_\ell$. While in the infinite-volume limit the squared amplitude is a rotationally symmetric function of the lepton momentum $\pvec_\ell$, this is not true in a finite volume. In fact, due to the breaking of rotational symmetry in a finite volume, finite-size effects can appear that depend on the direction of the lepton $\hat{\pvec}_\ell$. See ref.~\cite{Davoudi:2018qpl} for a detailed discussion of rotational symmetry breaking effects and refs.~\cite{Lubicz:2016xro,Tantalo:2016vxk,DiCarlo:2021apt} for their numerical evaluation in the context of leptonic decays. In this section we discuss the derivation of finite-volume effects to the leptonic decay rate in $\qedr$ up to $\mathrm{O}(1/L^3)$ and will show that, together with terms proportional to $\bar{c}_0$, which vanish in this regularization, momentum-dependent finite-volume effects also appear at $\mathrm{O}(1/L^3)$ multiplying a combination of known form factors.
Following ref.~\cite{DiCarlo:2021apt}, we write the finite-volume decay rate including corrections from a virtual photon as
\begin{equation}
    \Gamma_0(L) = \Gamma_0^\mathrm{tree}\bigg[1+2 \, \frac{\alpha_\mathrm{em}}{4\pi} \, Y(L)\bigg]\,,
\end{equation}
with $\Gamma_0^\mathrm{tree}$ denoting the tree-level decay rate. This process is diagrammatically depicted in \cref{fig:pl2diagram}.
Our aim is to determine the volume dependence of $Y(L)$. In order to do so by means of a large $L$ expansion of sum-integral differences, it is convenient to introduce an additional infrared regulator, like e.g. a photon mass $\lambda$, to regularize the infrared divergent infinite-volume integral and define
\begin{equation}
    Y(L) = \lim_{\lambda\to 0} \big[\Delta Y(L,\lambda) + Y_\mathrm{IV}^\mathrm{uni}(\lambda)\big]\,.
\end{equation}
The quantity $Y_\mathrm{IV}^\mathrm{uni}(\lambda)$ has been computed in perturbation theory in ref.~\cite{Lubicz:2016xro}, while $\Delta Y(L,\lambda)$ has been derived in ref.~\cite{DiCarlo:2021apt} in $\qedl$ and up to $\mathrm{O}(1/L^2)$, including a pointlike contribution at $\mathrm{O}(1/L^3)$. Here we derive $\Delta Y(L,\lambda)$ in $\qedr$ fully including the correction of $\mathrm{O}(1/L^3)$.
The squared decay amplitude can be separated in two contributions, depending on whether only the decaying hadron in \cref{fig:pl2diagram} interacts with the photon or the photon is exchanged between the hadron and the final state charged lepton.
We refer to the two contributions as factorizable and non-factorizable, respectively, and write accordingly
\begin{equation}
    |\mathcal{M}|^2 = |\mathcal{M}_0|^2 + |\mathcal{M}|_\text{fact.}^2 + |\mathcal{M}|_\text{non-fact.}^2\,,
\end{equation}
where the first term correspond to the squared amplitude with no photon corrections.
Then, the finite-volume correction $\Delta Y(L,\lambda)$ can be expressed in terms of sum-integral differences of the QED-corrected squared amplitude as
\begin{equation}
    2 \, \frac{\alpha_\mathrm{em}}{4\pi}\, \Delta Y(L,\lambda) = |\mathcal{M}_0|^{-2} \Big(\Delta|\mathcal{M}|_\text{fact.}^2 + \Delta|\mathcal{M}|_\text{non{-}fact.}^2\Big)\,,
\end{equation}
with
\begin{align}
    \Delta|\mathcal{M}|_\text{fact.}^2 &= \lim_{p^2\to-m_P^2} \Delta_\kvec^\mathrm{r} \int\frac{\dd k_0}{2\pi} \frac{1}{k^2+\lambda^2} \, H_W^{\rho\mu\mu}(p,k,-k) T_{\rho}(p,p_\ell)\,,
    \label{eq:corr_fact}
    \\
    \Delta|\mathcal{M}|_\text{non-fact.}^2 &= \lim_{p^2\to-m_P^2} \Delta_\kvec^\mathrm{r} \int\frac{\dd k_0}{2\pi} \frac{1}{k^2+\lambda^2} \, H_W^{\rho\mu}(p,k) T_{\rho\mu}(p,p_\ell,k)\,.
    \label{eq:corr_nonfact}
\end{align}
%%%
\begin{figure}[t!]
    \centering
    \begin{subfigure}{0.45\textwidth}
    \centering
        \includegraphics[width=0.95\linewidth]{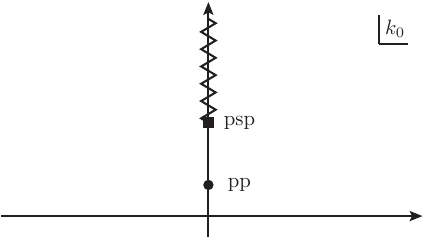}
        \caption{}
    \end{subfigure}%
      \begin{subfigure}{0.45\textwidth}
    \centering
        \includegraphics[width=0.95\linewidth]{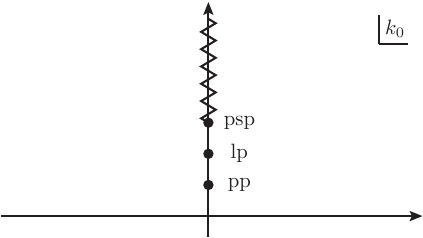}
        \caption{}
    \end{subfigure}%
    \caption{The analytic structure of the $k_0$ integrand for (a) the factorisable contribution in~\cref{eq:corr_fact}, and (b) the non-factorisable contribution in~\cref{eq:corr_nonfact}. The poles have been labeled pp (photon pole), psp (pseudoscalar pole) and lp (lepton pole). Single poles are denoted by regular circles, double poles by squares and branch cuts by zigzag lines. The positions of the singularities in general depend on $\mathbf{k}$. }
    \label{fig:lepdec_analytick0}
    \end{figure}
%%%
The quantities $T_{\rho}(p,p_\ell)$ and $T_{\rho\mu}(p,p_\ell,k)$ are leptonic tensors which depend on the lepton momentum $p_\ell=(i\omega_\ell(\pvec_\ell),\pvec_\ell)$ and are defined in ref.~\cite{DiCarlo:2021apt}. Here and in the following we use the notation $\omega_n(\pvec)=\sqrt{\pvec^2+m_n^2}$ to denote the energy of the particle $n$. Explicit expressions for the hadronic correlators $H_W^{\rho\mu\mu}(p,k,-k)$ and $H_W^{\rho\mu}(p,k)$ in terms of irreducible vertex functions can be obtained from ref.~\cite{DiCarlo:2021apt}.\footnote{The relation between the correlators used here and those studied in ref.~\cite{DiCarlo:2021apt} is $H_W^{\rho\mu\mu}(p,k,-k)=Z_P^{-1}D(p)^{-1} \, C_W^{\rho\mu\mu}(p,k,-k)$ and $H_W^{\rho\mu}(p,k) = Z_P^{-1}D(p)^{-1} \, C_W^{\rho\mu}(p,k)$.} In order to derive the finite-volume effects, we follow the same approach used in the previous section: we first perform the $k_0$ integral deforming the integration contour and encircling the singularities of the integrand, and then we expand the resulting expression for large value of $L$ after the substitution $\kvec=2\pi\nvec/L$. 
The analytical structure of the integrands in~\cref{eq:corr_fact,eq:corr_nonfact} is more complicated compared to the case of the Compton tensor, and shown in~\cref{fig:lepdec_analytick0}. In~\cref{eq:corr_fact}, in addition to the photon propagator pole in the upper half of the $k_0$ plane, the hadronic correlator has a double pole at $k_0=-p_0+i \omega_P(\kvec)$ corresponding to an on-shell intermediate meson, and a branch cut from the contribution of intermediate multiparticle states, as in the case of the Compton tensor. On the other hand, the non-factorizable correction in~\cref{eq:corr_nonfact} has, in addition to the photon pole in the upper $k_0$ plane, a single pole at $k_0=-p_0+i \omega_P(\kvec)$ coming from the hadronic correlator, a pole from the lepton propagator at $k_0=-p_{\ell,0}+i\omega_\ell(\pvec_\ell+\kvec)$, and a multi-particle branch cut. In~\cref{fig:lepdec_analytick0} the poles are shown with $p_0 = im_P$ and $p_{\ell, 0} = i \omega _\ell (\mathbf{p}_\ell)$. 

Applying the procedure described above and defining
\begin{equation}
    \Delta Y(L,\lambda) = \overline{Y}_\mathrm{log} \, \log\frac{L\lambda}{2\pi} + \sum_{n=0}^\infty \frac{\overline{Y}_n}{{L}^n} \,,
\end{equation}
we obtain the following results for the corrections up to  $\mathrm{O}(1/L^3)$\,,
\begin{align}
\overline{Y}_\mathrm{log} &= 2(1-2 A_1(\mathbf{v}_\ell))\,,\\
\overline{Y}_{0} & = \frac{\overline{c}_3-2\, (\overline{c}_3(\vlvec)-B_{1}(\vlvec))}{2\pi} + 2\, (1-\log 2) \, ,
	\\ 
	\overline{Y}_{1} & = -\frac{(1+r_{\ell}^2)^2\overline{c}_2-4\, r_{\ell}^2\overline{c}_{2}(\vlvec)}{m_{P}(1-r_{\ell}^4)} \, ,
	\\
	\overline{Y}_{2} & = -\frac{F_A^P}{f_P}\frac{4\pi \left[ (1+r_{\ell}^2)^2\overline{c}_1-4 \, r_{\ell}^2 \overline{c}_1(\vlvec) \right]}{m_P (1-r_{\ell}^4)}+\frac{8\pi\left[ (1+r_{\ell}^2)\overline{c}_1-2 \, \overline{c}_1 (\vlvec)\right] }{m_P^2 (1-r_{\ell}^4)} \, ,
	\\
    \overline{Y}_3  &= 
    \frac{\overline{c}_0}{m_P^3}\,  \frac{32\pi^2 \, (2+r_{\ell}^2)}{ (1+r_{\ell}^2)^3} + \bar{c}_0(\mathbf{v}_\ell) \, \mathcal{C}_\ell^\mathrm{(1)} + \bar{c}_0 \, \mathcal{C}_\ell^\mathrm{(2)} 
    \, ,
    \label{eq:Ybar_3}
\end{align}
where we have introduced the lepton velocity $\mathbf{v}_\ell=\pvec_\ell/\omega(\pvec_\ell)$ and the velocity-dependent finite volume-coefficients~\cite{Davoudi:2018qpl}
\begin{equation}
    \bar{c}_j(\mathbf{v}_\ell) = \Delta_\nvec^\mathrm{r} \, \frac{1}{|\nvec|^j}\frac{1}{(1-\mathbf{v}_\ell\cdot \hat{\nvec})}\, .
\end{equation}
The functions $A_1(\mathbf{v}_\ell)$ and $B_1(\mathbf{v}_\ell)$ are known functions defined in ref.~\cite{DiCarlo:2021apt}, while the quantity $F_A^P$ appearing in $\overline{Y}_2$ is the axial form factor associated to the amplitude of the real radiative decay $P\to\ell\bar{\nu}\gamma$. This form factor is a non-perturbative quantity that can be determined in lattice QCD\footnote{For the case when the photon in the radiative decay $P\rightarrow \ell \nu _\ell \gamma$ is virtual there are additional form factors. These can also be determined in e.g.~lattice QCD~\cite{Gagliardi:2022szw}.}~\cite{Desiderio:2020oej,Frezzotti:2020bfa,Giusti:2023pot,Gagliardi:2022szw,Frezzotti:2023ygt}, in chiral perturbation theory~\cite{Bijnens:1992en,Geng:2003mt} or measured in experiments~\cite{PhysRevLett.85.2256,PhysRevLett.89.061803,KLOE:2009urs,Tchikilev:2010wy,DUK201159,Cirigliano:2011ny}.
The quantity $\mathcal{C}_\ell^\mathrm{(1)}$ can be expressed in terms of known form factors
\begin{equation}
    \mathcal{C}_\ell^{(1)} = \frac{32\pi^2}{f_P \, m_P^2\, (1-r_\ell^4)}\, \left[ {\color{black}F_V^P }-{\color{black}F_A^P} +  
    2\,  r_\ell ^2 \, \frac{\partial F_A^{P }}{\partial x_\gamma }
    \right] \,,
    \label{eq:Cl1}
\end{equation}
with $F_V^P$ denoting the vector form factor associated to the $P\to\ell\bar{\nu}\gamma$ decay amplitude and $x_\gamma = 2\,{p\cdot k} /m_P^2$,
while the term $\mathcal{C}_\ell^{(2)}$ contains all other structure-dependent contributions which arise from the residues at the poles as well as from the integrals over the branch cuts in both factorizable and non factorizable correlators. As anticipated, we observe that while all terms proportional to $\bar{c}_0$ vanish in $\qedr$, the term $\bar{c}_0(\mathbf{v}_\ell)\,\mathcal{C}_\ell^\mathrm{(1)}$ survives. This term would be present in $\qedl$ and likely also in local finite-volume formulations like $\qedc$. In addition to the form factors $F_V^P$ and $F_A^P$, also the slope of $F_A^P$ with respect to the parameter $x_\gamma$ can be  determined from lattice calculations, as done in refs.~\cite{Desiderio:2020oej,Giusti:2023pot} for the radiative decay of the pion, kaon as well as $D$ and $D_s$ mesons. The appearance of such terms is not totally unexpected. In fact, it is related to collinear divergent terms in the infinite-volume amplitude, which contribute as $\log(1-|\mathbf{v}_\ell|)\sim \log(m_\ell/m_P)$. Due to the breaking of rotational symmetry in a finite volume, the logarithmic collinear divergences reduce to the non trivial functions $\bar{c}_j(\mathbf{v}_\ell)$, which also depend on the direction $\hat{\mathbf{v}}_\ell$. As discussed in ref.~\cite{Davoudi:2018qpl}, the $\qedl$ velocity-dependent coefficients $c_j(\mathbf{v})$ can be written as
\begin{equation}
    c_j(\mathbf{v}) = -\frac{1}{2|\mathbf{v}|}\log\bigg[\frac{1-|\mathbf{v}|}{1+|\mathbf{v}|}\bigg]\,  {c}_j + f_j(\mathbf{v})\,,
    \label{eq:cjv_vel_dependence}
\end{equation}
with the function $f_j(\mathbf{v})$ containing the dependence on the direction $\hat{\mathbf{v}}$, which gets stronger as ${|\mathbf{v}|\to 1}$, and vanishing when averaged over the solid angle of $\mathbf{v}$, namely $\tfrac{1}{4\pi}\int \dd\Omega_{\mathbf{v}} f_j(\mathbf{v})=0$. As a consequence, the coefficients $\bar{c}_0(\mathbf{v})$ are proportional to $c_j$ up to rotational breaking effects. This property will be used later in~\cref{sec:strategies_c0vl}, where two numerical strategies are proposed to remove the finite-volume effects proportional to $\bar{c}_0(\mathbf{v}_\ell)$ without the need of computing the form factors in~\cref{eq:Cl1}.

Before concluding this section, it is worth examining the contribution $\mathcal{C}_\ell^{(2)}$ in $\overline{Y}_3$ in greater detail. This term receives structure-dependent contributions from three distinct sources. First, the residues of the integrand functions in~\cref{eq:corr_fact,eq:corr_nonfact} at their respective poles can produce terms proportional to $\bar{c}_0$, in addition to those proportional to $\bar{c}_0(\mathbf{v}_\ell)$, which are absorbed into $\mathcal{C}_\ell^{(2)}$. Second, the integral over the branch cut of the factorizable correlator in~\cref{eq:corr_fact}, with the contributions from the poles already removed, can only yield functions with non-negative powers of $|\mathbf{k}|$. Consequently, the $1/L^3$ contribution can only be proportional to $\bar{c}_0$. Finally, a similar argument applies to the non-factorizable correlator in~\cref{eq:corr_nonfact}. The absence of terms proportional to $c_0(\mathbf{v}_\ell)$ from the branch cut can be directly verified through the spectral decomposition of the non-factorizable correlator.
We can start from the following quantity
\begin{align}
    \mathcal{C}_\mathrm{nf}(k; p_\ell) &= \lim_{p^2\to-m_P^2} H_W^{\rho\mu}(p,k) T_{\rho\mu}(p,p_\ell,k)\,, 
\end{align}
with
\begin{equation}
    \lim_{p^2\to-m_P^2} H_W^{\rho\mu}(p,k) = i\int \dd^4 x \, e^{ik\cdot x} \langle 0 | \mathrm{T} \big\{J^\mu(x) J^\rho_W(0)\big\} | P,\boldsymbol{0}\rangle \,.
\end{equation}
Separating the two time orderings and integrating over $x$ gives the following spectral decomposition
\begin{align}
    \mathcal{C}_\mathrm{nf}(k; p_\ell) &= \int_{0}^\infty \dd\mu^2 
      \frac{\langle 0|J_W^\rho(0)\, L^3 \delta_{\hat{\mathbf{P}},\kvec}\delta(\mu^2-\hat{Q}^2)\, J^\mu(0) |P,\mathbf{0} \rangle }{k_0-i (\sqrt{\mu^2+\kvec^2}-m_P)} \, \frac{L_{\rho\mu}(p,p_\ell,k)}{(k_0+i\omega_\ell(\pvec_\ell))^2 + \omega_\ell(\pvec_\ell+\kvec)^2} 
   \\
   &-  \int_{0}^\infty \dd\mu^2 \frac{\langle0| J^\mu(0)\, L^3 \delta_{\hat{\mathbf{P}},-\kvec}\delta(\mu^2-\hat{Q}^2) \, J_W^\rho(0)|P,\mathbf{0} \rangle}{k_0+i \sqrt{\mu^2+\kvec^2}}    
    \, \frac{L_{\rho\mu}(p,p_\ell,k)}{(k_0+i\omega_\ell(\pvec_\ell))^2 + \omega_\ell(\pvec_\ell+\kvec)^2}  \,, \nn
\end{align}
where $\hat{Q}^2$ denotes the invariant-mass operator, such that $\hat{Q}^2=\hat{H}^2-\hat{\mathbf{P}}^2$, with $\hat{H}$ and $\hat{\mathbf{P}}$ being the Hamiltonian and the momentum operators, respectively. In the expression above we have also made the denominator of the lepton propagator in $T_{\rho\mu}(p, p_\ell, k)$ explicit and denoted the remaining part of the leptonic tensor as $L_{\rho\mu}(p, p_\ell, k)$.
We note that the branch cut in the upper half of the complex $k_0$ plane comes from the first time ordering. The lightest multi-particle state contributing to the cut is made of two pions in the case $P=\pi$ or a pion and a kaon when $P=K$. Therefore, the lowest value for the branch point is at $k_0 = i m_\pi$, when $\kvec=\boldsymbol{0}$ and $\mu^2=(2m_\pi)^2$ for $P=\pi$ or $\mu^2=(m_\pi+m_K)^2$ for $P=K$. Setting $\kvec=\boldsymbol{0}$ in $\mathcal{C}_\mathrm{nf}(k; p_\ell)$ then yields
\begin{align}
    \mathcal{C}_\mathrm{nf}(k_0,\boldsymbol{0}; p_\ell) &= \int_{0}^\infty \dd\mu^2 
      \frac{\langle 0|J_W^\rho(0)\, L^3 \delta_{\hat{\mathbf{P}},\boldsymbol{0}}\delta(\mu^2-\hat{H}^2)\, J^\mu(0) |P,\mathbf{0} \rangle }{k_0-i (\mu-m_P)} \, \frac{L_{\rho\mu}(p,p_\ell,k_0)}{k_0(k_0+2i\omega_\ell(\pvec_\ell))} 
   \\
   &-  \int_{0}^\infty \dd\mu^2 \frac{\langle0| J^\mu(0)\, L^3 \delta_{\hat{\mathbf{P}},\mathbf{0}}\delta(\mu^2-\hat{H}^2) \, J_W^\rho(0)|P,\mathbf{0} \rangle}{k_0+i \mu}    
    \, \frac{L_{\rho\mu}(p,p_\ell,k_0)}{k_0(k_0+2i\omega_\ell(\pvec_\ell))}  \,, \nn
\end{align}
Since the contribution of the lepton pole, the single-particle pole ($\mu=m_P$) and the photon pole at $k_0 = 0$ are separated from the integral over the branch cut, the integral of the function $\mathcal{C}_\mathrm{nf}(k_0, \boldsymbol{0}; p_\ell)/k_0^2$ in~\cref{eq:corr_nonfact} along the multi-particle cut yields a finite result. Consequently, the contribution of the branch cut to $\Delta |\mathcal{M}|_\text{non-fact.}^2$ will be proportional to $\bar{c}_0$.

\subsection{Strategies to remove $\bar{c}_0(\mathbf{v}_{\ell})$}
\label{sec:strategies_c0vl}
\begin{figure}[t!]
    \centering
    \begin{subfigure}[t]{0.45\textwidth}
        \centering
        \includegraphics[width=\textwidth]{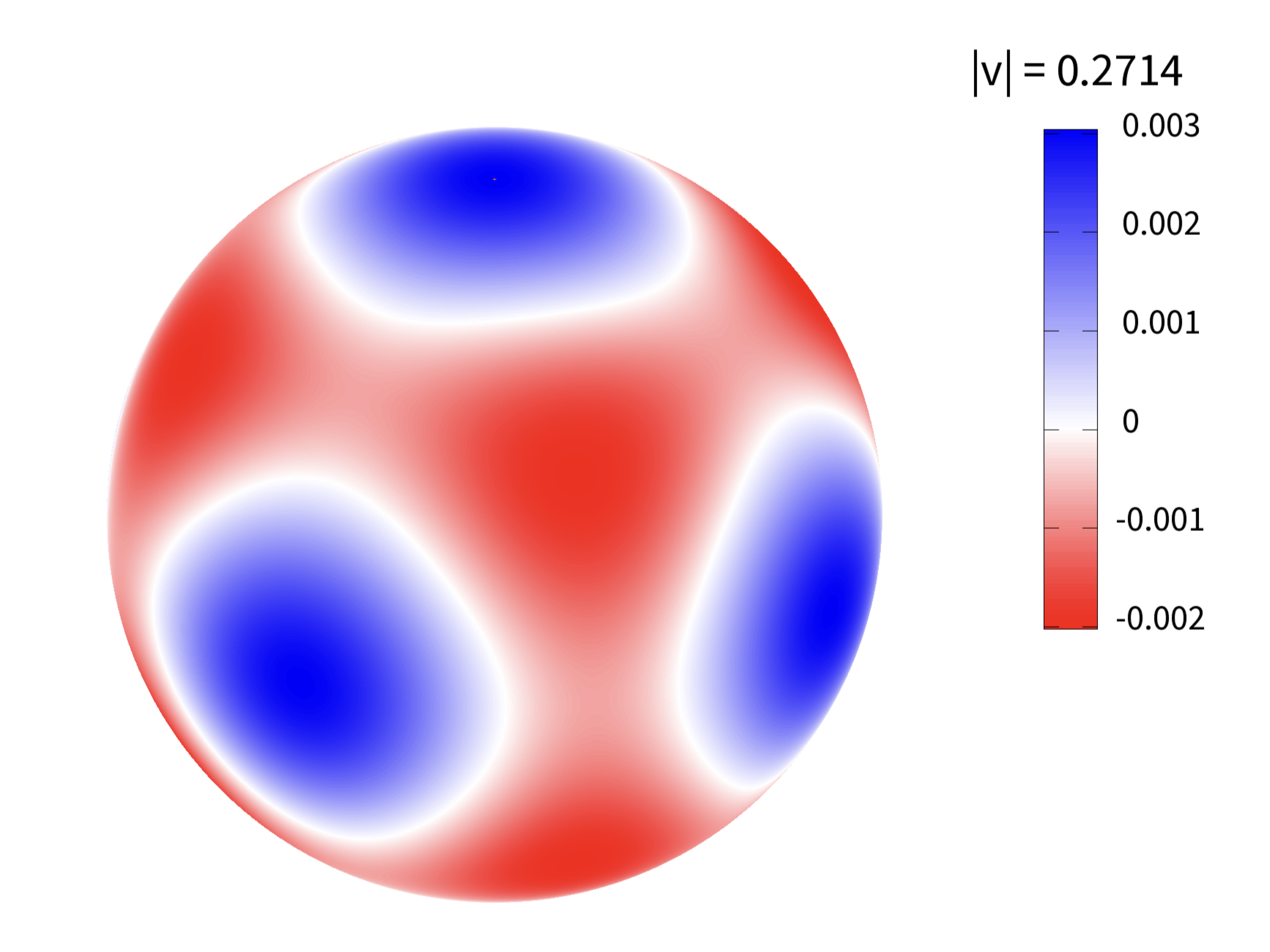}
        %\caption*{$|\mathbf{v}_\ell| = 0.2714$}
    \end{subfigure}
    \hfill
\begin{subfigure}[t]{0.45\textwidth}
        \centering
        \includegraphics[width=\textwidth]{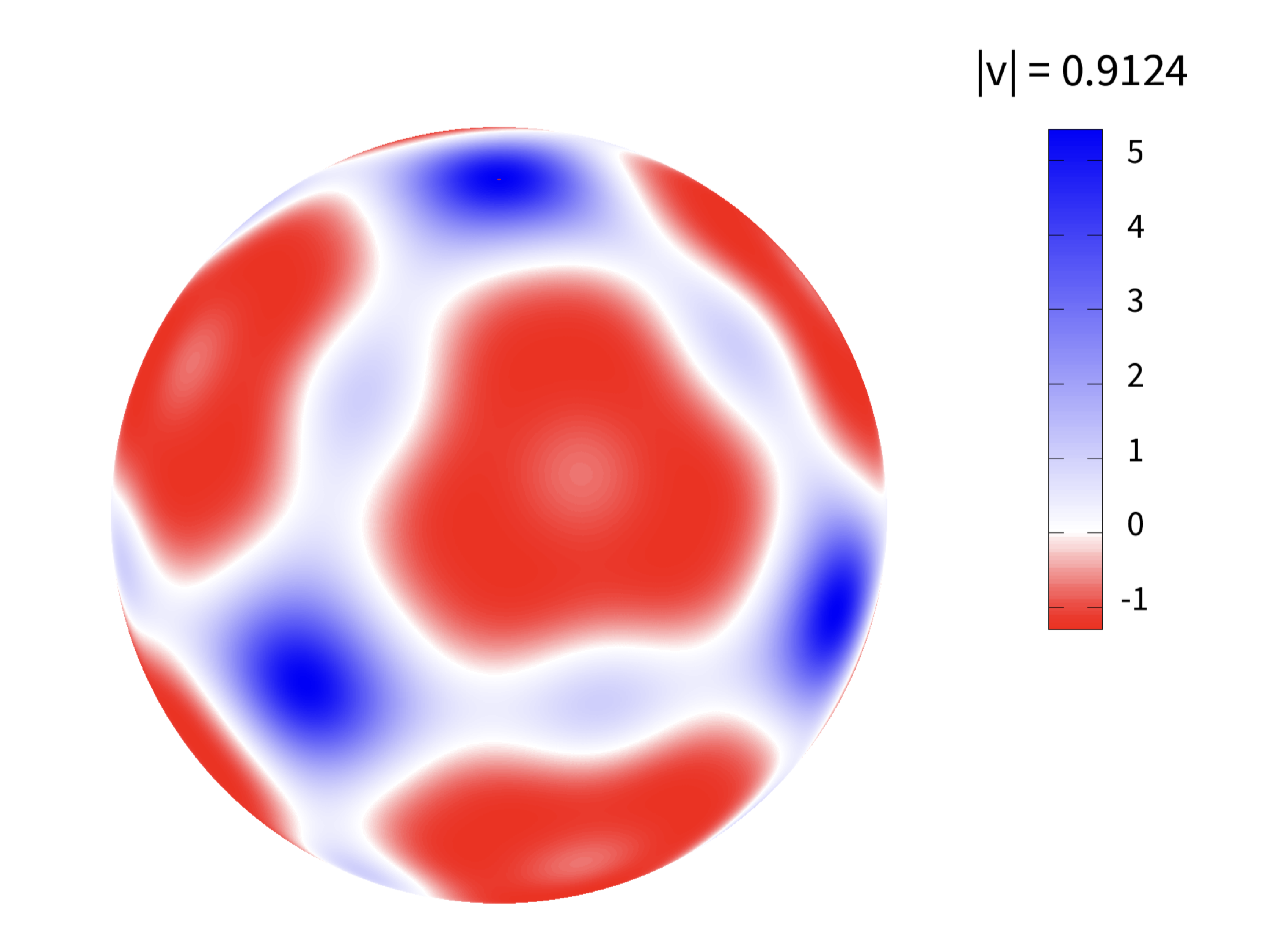}
        %\caption*{$|\mathbf{v}_\ell| = 0.9124$}
    \end{subfigure}

    \vspace{0.5em}

        \begin{subfigure}[t]{0.45\textwidth}
        \centering
        \includegraphics[width=\textwidth]{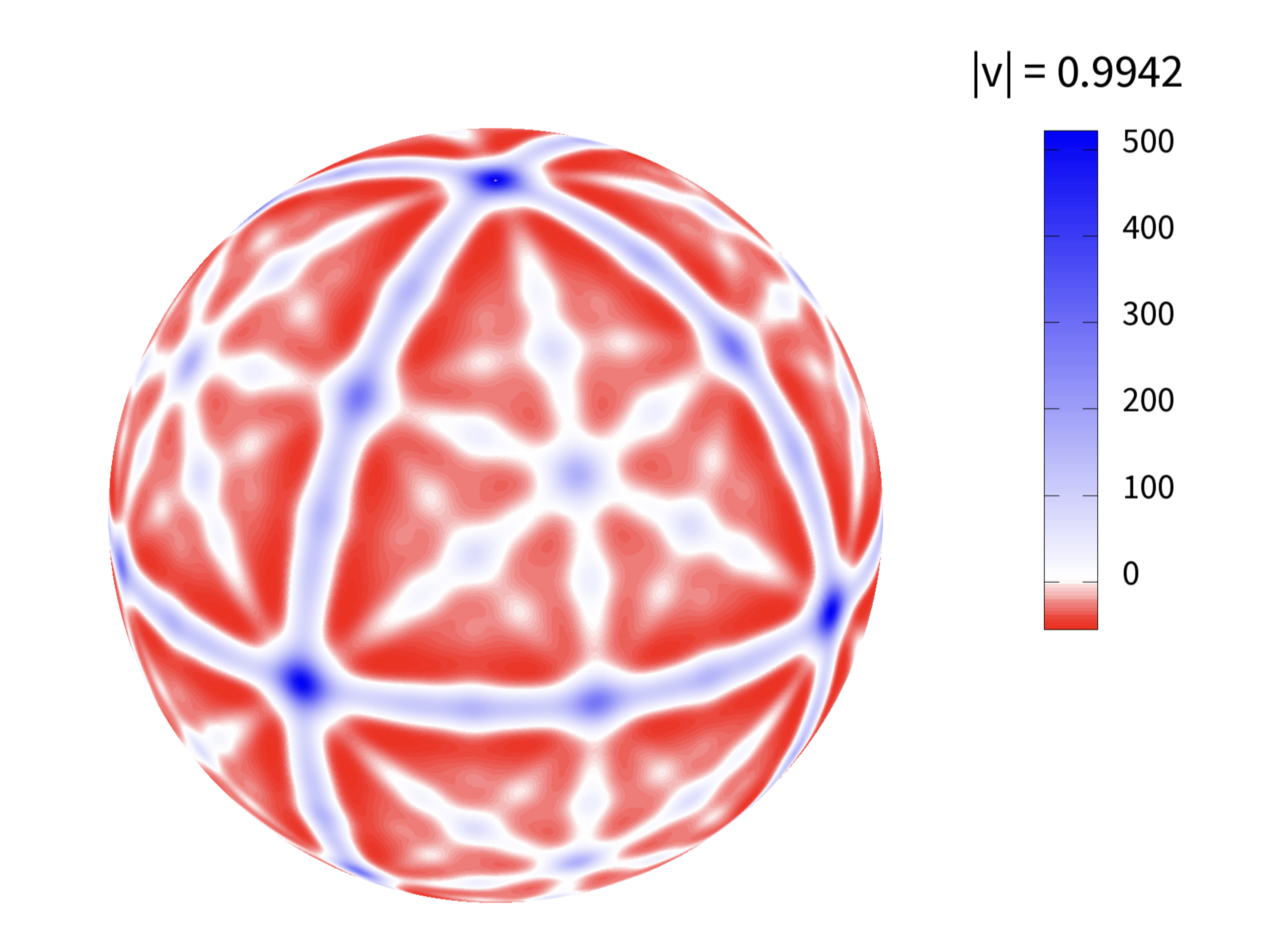}
        %\caption*{$|\mathbf{v}_\ell| = 0.9942$}
    \end{subfigure}
    \hfill
    \begin{subfigure}[t]{0.45\textwidth}
        \centering
        \includegraphics[width=\textwidth]{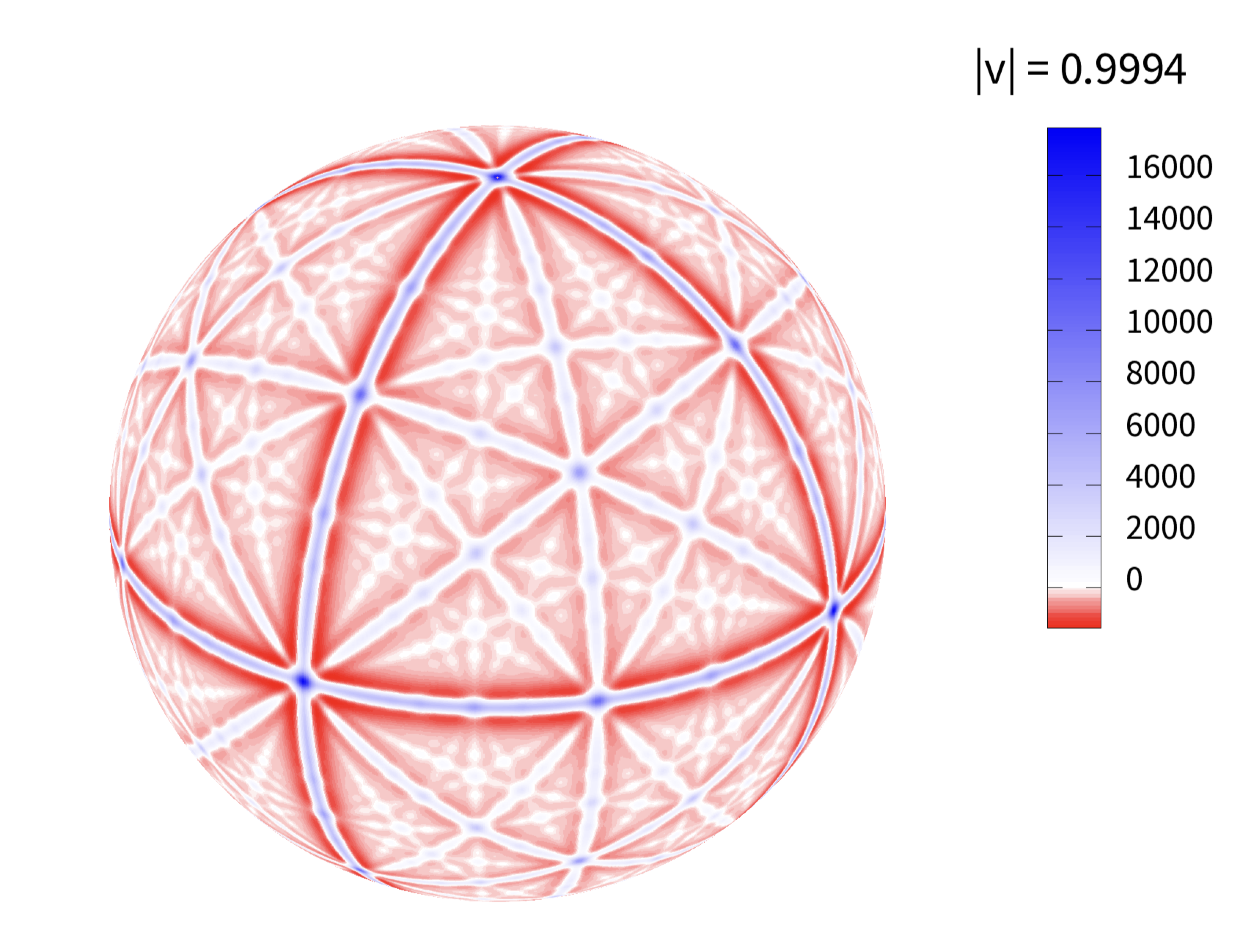}
        %\caption*{$|\mathbf{v}_\ell| = 0.9994$}
    \end{subfigure}

    \caption{Angular dependence of $\bar{c}_0(\mathbf{v}_\ell)$ for different values of $|\mathbf{v}_\ell|$. Positive and negative values are shown in blue and red, respectively. White regions correspond to directions for which $\bar{c}_0(\mathbf{v}_\ell)=0$.}
    \label{fig:c0angdep}
\end{figure}

While the coefficient $\bar{c}_{0}(\mathbf{v}_\ell)$, and thus $\overline{Y}_3/L^3$ in~\cref{eq:Ybar_3}, is generally non-zero in $\qedr$, it is still possible to device strategies to eliminate this contribution as well. This can be achieved by studying the dependence of $\bar{c}_{0}(\mathbf{v}_\ell)$ from the direction of the vector $\mathbf{v}_\ell$.
This can be written as
\begin{equation}
    \bar{c}_0(\mathbf{v}_\ell) = c_0(\mathbf{v}_\ell) + \sum_{\nvec} \frac{h(\nvec)}{1-\mathbf{v}_\ell\cdot \hat{\nvec}}\,,
\end{equation}
where the dependence on the velocity of the first term is given by~\cref{eq:cjv_vel_dependence} and the second term is such that
\begin{equation}
    \frac{1}{4\pi} \sum_\mathbf{n} \int \dd\Omega_{\mathbf{v}_\ell} \frac{h(\nvec)}{1-\mathbf{v}_\ell\cdot \hat{\nvec}} = -\frac{1}{2|\mathbf{v}_\ell|}\log\bigg[\frac{1-|\mathbf{v}_\ell|}{1+|\mathbf{v}_\ell|}\bigg] \, \sum_\mathbf{n} h(\nvec) = -\frac{1}{2|\mathbf{v}_\ell|}\log\bigg[\frac{1-|\mathbf{v}_\ell|}{1+|\mathbf{v}_\ell|}\bigg]\,.
\end{equation}
Combining this result with~\cref{eq:cjv_vel_dependence} and considering that $c_0=-1$, it follows that the $\qedr$ coefficient $\bar{c}_0(\mathbf{v}_\ell)$ is zero up to rotational breaking effects. The angular dependence of $\bar{c}_0(\mathbf{v}_\ell)$ is rather non trivial and it is shown in~\cref{fig:c0angdep} for four different values of the magnitude $|\mathbf{v}_\ell|$, namely $|\mathbf{v}_\ell| = \{0.2714,0.9124,0.9942,0.9994\}$. The values correspond to the velocity a muon would have in the leptonic decay of a pion, a kaon, a $D_s$ meson and a meson slightly heavier than the $B$, in their rest frame.
First, we observe that there are always directions for which $\bar{c}_0(\mathbf{v}_\ell)$ vanishes, regardless of the value of $|\mathbf{v}_\ell|$. These correspond to the white regions in the figure. Furthermore, as the magnitude $|\mathbf{v}_\ell|$ increases, a complex fractal pattern emerges, related to the number-theoretical properties of the components of the vector $\hat{\mathbf{v}}_\ell$. While the minimum value of $\bar{c}_0(\mathbf{v}_\ell)$ remains small in magnitude, its maximum diverges as $|\mathbf{v}_\ell| \to 1$. Simultaneously, however, the corresponding positive (blue) regions in~\cref{fig:c0angdep} become increasingly localized, balanced by larger negative (red) regions where the coefficient is negative. This provides a visual confirmation that the average of $\bar{c}_0(\mathbf{v}_\ell)$ over the solid angle must vanish.

From these observations, we identify two potential approaches to implement $\qedr$ improvement for momentum-dependent processes, such as the leptonic decays studied here. One approach, which has an easy implementation, consists in choosing the velocity of the lepton in the lattice calculation with a direction $\hat{\mathbf{v}}_{\ell}^\star$ such that $\bar{c}_0(\mathbf{v}_{\ell}^\star)=0$.\footnote{This approach can be readily implemented, facilitated by the \texttt{C++} code \texttt{QedFVCoef}~\cite{QedFvCoef}, which provides efficient evaluation of velocity-dependent coefficients through an auto-tuned algorithm based on that proposed in ref.~\cite{Davoudi:2018qpl} and extended in ref.~\cite{DiCarlo:2021apt}. Additionally, the code includes a Python binding equipped with a suite of convenient tools, such as a notebook that performs an angular scan and identifies directions for which $\bar{c}_0(\mathbf{v}_{\ell}^\star) = 0$.}

%%%%
%%%

A second approach leverages the fact the average of $\bar{c}_0(\mathbf{v}_\ell)$ over the direction of the velocity cancels, as discussed above. In a lattice calculation, this can be used by implementing a stochastic average of the lepton velocity directions, which are drawn randomly for each measurement. More explicitly, the average
\begin{align}
    \left\langle \bar{c}_0 (\mathbf{v}_{\ell}) \right\rangle = \frac{1}{N_{\text{meas}}}\, \sum _{i = 1}^{N_{\text{meas}}} \bar{c}_0 (\mathbf{v}_{\ell,\, i}) \, ,
\end{align}
where the $N_{\text{meas}}$ velocities $\mathbf{v}_{\ell,\, i}$ have a fixed norm $|\mathbf{v}_{\ell}|$ and directions randomly distributed according to the uniform distribution on the sphere;
 converges to zero in the limit of large statistics. This can be used to construct lattice Monte Carlo averages which do not suffer from finite-size effects proportional to $\bar{c}_0 (\mathbf{v}_{\ell})$. The stochastic averaging approach is potentially more computationally demanding than simply finding a velocity $\mathbf{v}_{\ell}^\star$ such that $\bar{c}_0(\mathbf{v}_{\ell}^\star) = 0$. However, its significant advantage lies in its ability to eliminate finite-volume effects that break rotational symmetry not only in the coefficient $\bar{c}_0(\mathbf{v}_\ell)$, but also in all other coefficients~$\bar{c}_j(\mathbf{v}_\ell)$.

\section{Conclusions}\label{sec:conclusions}
In this paper we have introduced a new finite-volume formulation of QED, $\qedr$, which is a specific choice within the framework of infrared improved $\qedl$ originally introduced in ref.~\cite{Davoudi:2018qpl}. It is designed to remove finite-volume effects at order $1/L^3$ in the large-volume expansion of observables, by changing the underlying QED action through the inclusion of a set of weights acting on a finite number of Fourier modes. The added weights alter the values of the finite-volume coefficients appearing in the large-volume expansion, and the chosen set of weights in $\qedr$ sets to zero $\bar{c}_0$ which appears at order $1/L^3$ for pseudoscalar masses and leptonic decay rates. For leptonic decay rates it was found that also the moment-dependent coefficient $\bar{c}_{0}(\mathbf{v})$ appears at order $1/L^3$, whose value depends on the direction of the momentum. It was shown that these remaining finite-volume effects also can be removed in $\qedr$ by either cleverly choosing the kinematics or stochastically averaging the lattice calculations over a range of different kinematics.

Our results are highly relevant for analytical correction of lattice data in precision calculations of e.g.~flavour physics observables. In ref.~\cite{Boyle:2022lsi} we studied isospin-breaking corrections to leptonic decays of pions and kaons, relevant for the extraction of the CKM elements $|V_{us}|$ and $|V_{ud}|$. We observed a major precision bottleneck due to not being able to predict the finite-volume effects at order $1/L^3$ in $\qedl$. Identically removing these contributions with a minor modification of the QED action is a major advantage of using $\qedr$ in modern lattice calculations. We are currently running QCD+QED simulations with $\qedr$ to supersede our results for leptonic decays in ref.~\cite{Boyle:2022lsi}. This will be presented in a future publication. Finally, independently of the QED prescription adopted, it remains essential to perform simulations at multiple volumes to enable a controlled extrapolation to the infinite-volume limit. 

\section*{Acknowledgements}

M.T.H., and A.P. are supported in part by UK STFC
grants ST/T000600/1 and ST/X000494/1, and M.D.C by UK STFC grant ST/P000630/1. M.T.H.~is additionally supported by UKRI Future Leader Fellowship
MR/T019956/1. M.D.C.~and A.P.~received funding from the European Research Council (ERC) under the European Union’s Horizon 2020 research
and innovation programme under grant agreement No 757646 and A.P.~additionally under
grant agreement No 813942. N.~H.-T.~was initially funded by the Swedish Research Council, project number 2021-06638, and currently by the UK Research and Innovation, Engineering and Physical Sciences Research Council, grant number EP/X021971/1.
M.D.C. has received funding from the European Union’s Horizon Europe research and innovation programme under the Marie Sklodowska-Curie grant agreement No 101108006.

\appendix

\section{Finite-volume coefficients in different QED formulations}\label{app:qedcfvcoeffs}
Here we study values of the finite-volume coefficients entering the large-volume expansions presented in this paper, elaborating on the results from the proceedings ref.~\cite{Hermansson-Truedsson:2023krp}. We present results for both $\qedl$ and $\qedr$, but in addition consider the analogous coefficients in $\qedc$ associated to the change of photon boundary conditions. It should be noted that in $\qedc$ also the fermion boundary conditions change, which in turn can affect the large-volume expansion. As a derivation of these finite-volume effects in $\qedc$ is outside the scope of this paper, we simply consider the $\qedc$ analogues of finite-volume coefficients associated to the photon boundary conditions. To study the three formulations we introduce the generalised sum-integral difference operator depending on the prescription $\textrm{QED}_{\textrm{X}}$ (where $\textrm{X}=\textrm{L}, \textrm{r}, \textrm{C}$)
\begin{align}
    \Delta _{\mathbf{n}}^{\textrm{X}} = 
    \left[
    \sum _{\mathbf{n}\in \Pi ^{\textrm{X}}} -\int \dd^3 \mathbf{n} 
    \right] \left(1+h^{\textrm{X}}(\mathbf{n})\right) \, .
\end{align}
The weights $h^{\textrm{L}}(\mathbf{n}) = h^{\textrm{C}}(\mathbf{n})= 0$ whereas $h^{\textrm{r}}(\mathbf{n}) = \delta _{|\mathbf{n}|,1}/6$. The sets of allowed momenta $\Pi ^{\textrm{X}}$ are
\begin{align}
  \Pi ^\textrm{L} = \Pi ^{\textrm{r}} & =  \left\{  
	 {\mathbf{n}}\in  \mathbb{Z}^3\setminus (0,0,0)
	\right\} \, ,
\\
    \Pi ^\textrm{C} & =  \left\{  
	 {\mathbf{n}} = \tilde{\mathbf{n}}+ 
     \frac{1}{2}\, (1,1,1)
     ,\, \tilde{\mathbf{n}} \in \mathbb{Z}^3
	\right\}\, .
\end{align}
As can be seen, from the point-of-view of the QED action and photon propagator the only difference between formulations is over which momentum modes $\mathbf{n}$ the sum runs. We can thus generalise the finite-volume coefficients $b_{j} (\mathbf{v} , \xi)$ from ref.~\cite{DiCarlo:2021apt} to
\begin{align}\label{eq:bjL}
	b_{j}^{\textrm{X}} (\mathbf{v} , \xi) = \Delta _{\mathbf{n}}^{\textrm{X}}\,  \left[ \frac{1}{\left[ \mathbf{n}^2 +\xi ^2\right] ^{j/2}\, ( 1-\hat{\mathbf{n}}_\xi \cdot \mathbf{v} ) }  \right] \, ,
\end{align}
where $\xi$ is an IR-regulator, $j$ is allowed to take any integer value and $\hat{\mathbf{n}}_\xi  =\hat{\mathbf{n}}\, |\mathbf{n}|  / \sqrt{\mathbf{n}^2+ \xi^2 }$.

The $b_j^{\textrm{X}} (\mathbf{v} , \xi)$ are related to the finite\footnote{In this appendix we have chosen the notation $c_j^{\textrm{X}} (\mathbf{v} )$ as it provides a unified way to express the coefficients across QED prescriptions. The relations to the $\qedl$ and $\qedr$ coefficients are $c_j^{\textrm{L}} (\mathbf{v} ) = c_j (\mathbf{v} )$ and $c_j^{\textrm{r}} (\mathbf{v} ) = \bar{c}_j (\mathbf{v} )$. 
} $c_j^{\textrm{X}} (\mathbf{v} )$ through
\begin{align}
&\underline{j<3:} \qquad 	
\hspace{0.3cm} {b}_{j}^{\textrm{X}} (\mathbf{v})
 =
{c}_{j}^{\textrm{X}}(\mathbf{v})=\Delta _{\mathbf{n}}^{\textrm{X}} \left[ \frac{1}{|{\mathbf{n}} | ^{j}\, ( 1-\hat{{\mathbf{n}}} \cdot \mathbf{v} )}  \right] \, ,
\\
&\underline{j=3:} \qquad 
{b} _{3}^{\textrm{X}} (\mathbf{v},\xi)
=
{c} _{3}^{\textrm{X}} (\mathbf{v})+4\pi \, A(\mathbf{v}) \, \log \xi-B(\mathbf{v})
\, , \\
& \hspace{2.05cm} {c}_3^{\textrm{X}} (\mathbf{v})
 =
\lim _{R\to\infty}\left[ \left. \sum _{\mathbf{n}\in \Pi^\textrm{X}: \, |{\mathbf{n}}| <R} \right.  \frac{1+h^{\textrm{X}}(\mathbf{n})}{|{\mathbf{n}}| ^{3}\, ( 1-\hat{{\mathbf{n}}} \cdot \mathbf{v} )} 
-4\pi \, A(\mathbf{v})\, \log R\right]
\, ,
\\
&\underline{j>3:} \qquad 
	{b} _{j}^{\textrm{X}}(\mathbf{v},\xi )
	=
	{c} _{j}^{\textrm{X}}(\mathbf{v})-\frac{\pi^{3/2}}{\xi^{j-3}}\frac{\Gamma \left( \frac{j-3}{2}\right)}{\Gamma (j)}\, F _{2,1}\left( \frac{1}{2},1,\frac{j}{2},\mathbf{v}^2\right)
	\, , \\
& \hspace{2.05cm} {c}_j^{\textrm{X}}(\mathbf{v})
	 =
\left. \sum _{\mathbf{n}\in \Pi ^{\textrm{X}}}\right. \left[ \frac{1+h^{\textrm{X}}(\mathbf{n})}{|{\mathbf{n}}| ^{j}\, ( 1-\hat{{\mathbf{n}}} \cdot \mathbf{v} )}  \right] 
\, .
\end{align}
Here $F _{2,1}$ is a hypergeometric function, $\Gamma$ is the Gamma function, and $A(\mathbf{v})$ as well as $B(\mathbf{v})$ are known integrals given in ref.~\cite{DiCarlo:2021apt}. 

Matching the $c_j^{\textrm{C}}$ to the $\xi_{j}$ appearing for the hadron masses in ref.~\cite{Lucini:2015hfa}, which are related to generalised zeta function, one immediately finds the relations
\begin{align}\label{eq:qedcmatchingc1}
	c_{1}^{\textrm{C}} & =\frac{\xi_{2}}{\pi} \, ,
	\\
	c_{2}^{\textrm{C}} & =\pi \xi_{1} 
	\, .
	\label{eq:qedcmatchingc2}
\end{align}
This result is numerically validated below. We stress that a matching for leptonic decays cannot be done, as it is currently unknown how to extract the isospin-breaking corrections to leptonic decays in $\qedc$, meaning that the function $Y(L)$ studied in this paper might not be the relevant finite-volume function. 

\subsection{Numerical comparison between $\qedr$, $\qedl$ and $\qedc$ }
Next we use the numerical algorithm developed in refs.~\cite{Davoudi:2018qpl,DiCarlo:2021apt} to evaluate the finite-volume coefficients ${c}_{j}^{\textrm{X}}(\mathbf{v})$ in $\qedl$, $\qedr$ and $\qedc$. Again, the only difference in employing the algorithm comes from the allowed momenta in $\Pi^{\textrm{X}}$. We here focus on the coefficients appearing for the self-energy and leptonic decays in $\qedl$ and $\qedr$, and their equivalents in $\qedc$.
The velocity-dependent coefficients depend on the orientation of the velocity due to the breaking of rotational symmetry in the finite volume. Here we consider the velocity $\mathbf{v} = \frac{|\mathbf{v} |}{\sqrt{3}} (1,1,1)$, where $|\mathbf{v}|=0.912401 $ corresponds to the velocity the final-state muon in the physical decay $K\rightarrow \mu \nu _\mu$ in the kaon rest frame. The norm depends on the kaon and muon masses through $|\mathbf{v}| = [1-(m_K/m_\mu)^2]/[1+(m_K/m_\mu)^2]$, with PDG values~\cite{ParticleDataGroup:2024cfk} $m_K = 0.493677$ GeV and $m_\mu = 0.10565837$ GeV. We also include the zero-velocity limit of the coefficients. 

The values are presented in~\cref{table:cjv}. As a cross-check, the values of $	c_{1}^{\textrm{C}}$ and $	c_{2}^{\textrm{C}}$ satisfy the matching~\cref{eq:qedcmatchingc1,eq:qedcmatchingc2} to ref.~\cite{Lucini:2015hfa} as expected. Note that for $j=0,3$ the $\qedl$ and $\qedr$ values in magnitude are smaller than those in $\qedc$. The contrary is true for $j=1,2$. By varying the norm $|\mathbf{v}|$ (not shown in the table) one observes that the hierarchies can change. For instance, calculating $j=0$ for a range of $|\mathbf{v}|$ shows that the magnitude of $c_0^{\textrm{C}} (\mathbf{v})$ becomes smaller than $c_0^{\textrm{L}} (\mathbf{v})$ around $|\mathbf{v}|\approx 0.87$. We stress that unlike for $c_0^{\textrm{C}} $, the locality of $\qedc$ does not set to zero the $c_0^{\textrm{C}}(\mathbf{v}) $ and its numerical value is in general as sizeable as in $\qedl$ and $\qedr$. We again emphasise that large-volume expansions do not necessarily take the same functional form in $\qedc$ as compared to $\qedl$ and $\qedr$. The actual comparison of finite-volume effects between different QED formulations requires the knowledge of all the relevant volume expansions, which goes beyond the scope of this work.

\begin{table}[t!]
	\centering
	\begin{tabular}{|c|c|c|c|c|c|c|}
		\hline
		$j$  & 
		{\color{black}$c^{\textrm{L}}_{j}(\mathbf{v})$} & {\color{black}${c}^{\textrm{r}}_{j}(\mathbf{v})$} & {\color{black}$c^{\textrm
  C} _{j}(\mathbf{v})$} & {\color{black}$c_{j}^{\textrm{L}}$} & {\color{black}$c^{\textrm{r}}_{j}$} &{\color{black}$c^{\textrm{C}} _{j}$} \\
		\hline
  3  & 
 4.9451 & 6.3292 & 22.4744 &  3.8219 & 4.8219 & 8.9655
  \\
		2  & 
		-16.3454 & -14.9613 &-3.2067 & -8.9136 & -7.9136 &-5.4901 
		\\
		1   & 
		-5.7302 & -4.3461 & 3.5122 & -2.8373 & -1.8373 &-0.8019 
		\\
		\color{black}0  &
		\color{black} -2.1237 & \color{black}-0.7396 &	\color{black} 3.6927 & 	\color{black} -1 &  {\color{black}0} &	\color{black} 0 
		\\
		\hline
	\end{tabular}
		\caption{Finite-volume coefficients ${c}^{\textrm{L}}_{j}\left(\mathbf{v}\right)$, ${c}^{\textrm{r}}_j (\mathbf{v})$ and ${c}^{\textrm{C}}_{j}\left(\mathbf{v}\right)$, in $\qedl$, $\qedr$ and $\qedc$, respectively. }
		\label{table:cjv}
\end{table}

%References
\bibliography{refs}

\providecommand{\href}[2]{#2}\begingroup\raggedright\begin{thebibliography}{10}

\bibitem{FlavourLatticeAveragingGroupFLAG:2024oxs}
{\scshape Flavour Lattice Averaging Group (FLAG)} collaboration, \emph{{FLAG
  Review 2024}},  \href{https://arxiv.org/abs/2411.04268}{{\ttfamily
  2411.04268}}.

\bibitem{deDivitiis:2013xla}
{\scshape RM123} collaboration, \emph{{Leading isospin breaking effects on the
  lattice}}, \href{https://doi.org/10.1103/PhysRevD.87.114505}{\emph{Phys. Rev.
  D} {\bfseries 87} (2013) 114505}
  [\href{https://arxiv.org/abs/1303.4896}{{\ttfamily 1303.4896}}].

\bibitem{BMW:2014pzb}
{\scshape BMW} collaboration, \emph{{Ab initio calculation of the
  neutron-proton mass difference}},
  \href{https://doi.org/10.1126/science.1257050}{\emph{Science} {\bfseries 347}
  (2015) 1452} [\href{https://arxiv.org/abs/1406.4088}{{\ttfamily 1406.4088}}].

\bibitem{Clark:2022wjy}
M.A.~Clark, M.~Della~Morte, Z.~Hall, B.~H\"orz, A.~Nicholson, A.~Shindler
  et~al., \emph{{QED with massive photons for precision physics: zero modes and
  first result for the hadron spectrum}},
  \href{https://doi.org/10.22323/1.396.0281}{\emph{PoS} {\bfseries LATTICE2021}
  (2022) 281} [\href{https://arxiv.org/abs/2201.03251}{{\ttfamily
  2201.03251}}].

\bibitem{Frezzotti:2022dwn}
R.~Frezzotti, G.~Gagliardi, V.~Lubicz, G.~Martinelli, F.~Sanfilippo and
  S.~Simula, \emph{{Lattice calculation of the pion mass difference
  M\ensuremath{\pi}+-M\ensuremath{\pi}0 at order O(\ensuremath{\alpha}em)}},
  \href{https://doi.org/10.1103/PhysRevD.106.014502}{\emph{Phys. Rev. D}
  {\bfseries 106} (2022) 014502}
  [\href{https://arxiv.org/abs/2202.11970}{{\ttfamily 2202.11970}}].

\bibitem{RCstar:2022yjz}
{\scshape RCstar} collaboration, \emph{{First results on QCD+QED with C$^{*}$
  boundary conditions}},
  \href{https://doi.org/10.1007/JHEP03(2023)012}{\emph{JHEP} {\bfseries 03}
  (2023) 012} [\href{https://arxiv.org/abs/2209.13183}{{\ttfamily
  2209.13183}}].

\bibitem{Segner:2023igh}
A.M.~Segner, A.~Risch and H.~Wittig, \emph{{Precision Determination of Baryon
  Masses including Isospin-breaking}},
  \href{https://doi.org/10.22323/1.453.0044}{\emph{PoS} {\bfseries LATTICE2023}
  (2024) 044} [\href{https://arxiv.org/abs/2312.09065}{{\ttfamily
  2312.09065}}].

\bibitem{Giusti:2017jof}
D.~Giusti, V.~Lubicz, G.~Martinelli, F.~Sanfilippo and S.~Simula,
  \emph{{Strange and charm HVP contributions to the muon ($g - 2)$ including
  QED corrections with twisted-mass fermions}},
  \href{https://doi.org/10.1007/JHEP10(2017)157}{\emph{JHEP} {\bfseries 10}
  (2017) 157} [\href{https://arxiv.org/abs/1707.03019}{{\ttfamily
  1707.03019}}].

\bibitem{RBC:2018dos}
{\scshape RBC, UKQCD} collaboration, \emph{{Calculation of the hadronic vacuum
  polarization contribution to the muon anomalous magnetic moment}},
  \href{https://doi.org/10.1103/PhysRevLett.121.022003}{\emph{Phys. Rev. Lett.}
  {\bfseries 121} (2018) 022003}
  [\href{https://arxiv.org/abs/1801.07224}{{\ttfamily 1801.07224}}].

\bibitem{Giusti:2019xct}
D.~Giusti, V.~Lubicz, G.~Martinelli, F.~Sanfilippo and S.~Simula,
  \emph{{Electromagnetic and strong isospin-breaking corrections to the muon $g
  - 2$ from Lattice QCD+QED}},
  \href{https://doi.org/10.1103/PhysRevD.99.114502}{\emph{Phys. Rev. D}
  {\bfseries 99} (2019) 114502}
  [\href{https://arxiv.org/abs/1901.10462}{{\ttfamily 1901.10462}}].

\bibitem{Aoyama:2020ynm}
T.~Aoyama et~al., \emph{{The anomalous magnetic moment of the muon in the
  Standard Model}},
  \href{https://doi.org/10.1016/j.physrep.2020.07.006}{\emph{Phys. Rept.}
  {\bfseries 887} (2020) 1} [\href{https://arxiv.org/abs/2006.04822}{{\ttfamily
  2006.04822}}].

\bibitem{Borsanyi:2020mff}
S.~Borsanyi et~al., \emph{{Leading hadronic contribution to the muon magnetic
  moment from lattice QCD}},
  \href{https://doi.org/10.1038/s41586-021-03418-1}{\emph{Nature} {\bfseries
  593} (2021) 51} [\href{https://arxiv.org/abs/2002.12347}{{\ttfamily
  2002.12347}}].

\bibitem{Altherr:2022fqa}
A.~Altherr et~al., \emph{{Hadronic vacuum polarization with C* boundary
  conditions}}, \href{https://doi.org/10.22323/1.430.0312}{\emph{PoS}
  {\bfseries LATTICE2022} (2023) 312}
  [\href{https://arxiv.org/abs/2212.11551}{{\ttfamily 2212.11551}}].

\bibitem{Biloshytskyi:2022ets}
V.~Biloshytskyi, E.-H.~Chao, A.~G\'erardin, J.R.~Green, F.~Hagelstein,
  H.B.~Meyer et~al., \emph{{Forward light-by-light scattering and
  electromagnetic correction to hadronic vacuum polarization}},
  \href{https://doi.org/10.1007/JHEP03(2023)194}{\emph{JHEP} {\bfseries 03}
  (2023) 194} [\href{https://arxiv.org/abs/2209.02149}{{\ttfamily
  2209.02149}}].

\bibitem{Chao:2023lxw}
E.-H.~Chao, H.B.~Meyer and J.~Parrino, \emph{{Coordinate-space calculation of
  QED corrections to the hadronic vacuum polarization contribution to
  $(g-2)_\mu$}}, \href{https://doi.org/10.22323/1.453.0256}{\emph{PoS}
  {\bfseries LATTICE2023} (2024) 256}
  [\href{https://arxiv.org/abs/2310.20556}{{\ttfamily 2310.20556}}].

\bibitem{Boccaletti:2024guq}
A.~Boccaletti et~al., \emph{{High precision calculation of the hadronic vacuum
  polarisation contribution to the muon anomaly}},
  \href{https://arxiv.org/abs/2407.10913}{{\ttfamily 2407.10913}}.

\bibitem{Djukanovic:2024cmq}
D.~Djukanovic, G.~von Hippel, S.~Kuberski, H.B.~Meyer, N.~Miller, K.~Ottnad
  et~al., \emph{{The hadronic vacuum polarization contribution to the muon
  $g-2$ at long distances}},
  \href{https://arxiv.org/abs/2411.07969}{{\ttfamily 2411.07969}}.

\bibitem{Parrino:2025afq}
J.~Parrino, V.~Biloshytskyi, E.-H.~Chao, H.B.~Meyer and V.~Pascalutsa,
  \emph{{Computing the UV-finite electromagnetic corrections to the hadronic
  vacuum polarization in the muon $(g-2)$ from lattice QCD}},
  \href{https://arxiv.org/abs/2501.03192}{{\ttfamily 2501.03192}}.

\bibitem{Carrasco:2015xwa}
N.~Carrasco, V.~Lubicz, G.~Martinelli, C.T.~Sachrajda, N.~Tantalo, C.~Tarantino
  et~al., \emph{{QED Corrections to Hadronic Processes in Lattice QCD}},
  \href{https://doi.org/10.1103/PhysRevD.91.074506}{\emph{Phys. Rev. D}
  {\bfseries 91} (2015) 074506}
  [\href{https://arxiv.org/abs/1502.00257}{{\ttfamily 1502.00257}}].

\bibitem{Giusti:2017dwk}
D.~Giusti, V.~Lubicz, G.~Martinelli, C.T.~Sachrajda, F.~Sanfilippo, S.~Simula
  et~al., \emph{{First lattice calculation of the QED corrections to leptonic
  decay rates}},
  \href{https://doi.org/10.1103/PhysRevLett.120.072001}{\emph{Phys. Rev. Lett.}
  {\bfseries 120} (2018) 072001}
  [\href{https://arxiv.org/abs/1711.06537}{{\ttfamily 1711.06537}}].

\bibitem{DiCarlo:2019thl}
M.~Di~Carlo, D.~Giusti, V.~Lubicz, G.~Martinelli, C.T.~Sachrajda, F.~Sanfilippo
  et~al., \emph{{Light-meson leptonic decay rates in lattice QCD+QED}},
  \href{https://doi.org/10.1103/PhysRevD.100.034514}{\emph{Phys. Rev. D}
  {\bfseries 100} (2019) 034514}
  [\href{https://arxiv.org/abs/1904.08731}{{\ttfamily 1904.08731}}].

\bibitem{Boyle:2022lsi}
P.~Boyle et~al., \emph{{Isospin-breaking corrections to light-meson leptonic
  decays from lattice simulations at physical quark masses}},
  \href{https://doi.org/10.1007/JHEP02(2023)242}{\emph{JHEP} {\bfseries 02}
  (2023) 242} [\href{https://arxiv.org/abs/2211.12865}{{\ttfamily
  2211.12865}}].

\bibitem{Christ:2023lcc}
N.H.~Christ, X.~Feng, L.-C.~Jin, C.T.~Sachrajda and T.~Wang, \emph{{Radiative
  corrections to leptonic decays using infinite-volume reconstruction}},
  \href{https://doi.org/10.1103/PhysRevD.108.014501}{\emph{Phys. Rev. D}
  {\bfseries 108} (2023) 014501}
  [\href{https://arxiv.org/abs/2304.08026}{{\ttfamily 2304.08026}}].

\bibitem{Kronfeld:1990qu}
A.S.~Kronfeld and U.J.~Wiese, \emph{{SU(N) gauge theories with C periodic
  boundary conditions. 1. Topological structure}},
  \href{https://doi.org/10.1016/0550-3213(91)90479-H}{\emph{Nucl. Phys. B}
  {\bfseries 357} (1991) 521}.

\bibitem{Wiese:1991ku}
U.J.~Wiese, \emph{{C periodic and G periodic QCD at finite temperature}},
  \href{https://doi.org/10.1016/0550-3213(92)90333-7}{\emph{Nucl. Phys. B}
  {\bfseries 375} (1992) 45}.

\bibitem{Kronfeld:1992ae}
A.S.~Kronfeld and U.J.~Wiese, \emph{{SU(N) gauge theories with C periodic
  boundary conditions. 2. Small volume dynamics}},
  \href{https://doi.org/10.1016/0550-3213(93)90302-6}{\emph{Nucl. Phys. B}
  {\bfseries 401} (1993) 190}
  [\href{https://arxiv.org/abs/hep-lat/9210008}{{\ttfamily hep-lat/9210008}}].

\bibitem{Polley:1993bn}
L.~Polley, \emph{{Boundaries for $SU(3)_C \times U(1)_{\textrm{el}}$ lattice
  gauge theory with a chemical potential}},
  \href{https://doi.org/10.1007/BF01555844}{\emph{Z. Phys. C} {\bfseries 59}
  (1993) 105}.

\bibitem{Duncan:1996be}
A.~Duncan, E.~Eichten and H.~Thacker, \emph{{Electromagnetic structure of light
  baryons in lattice QCD}},
  \href{https://doi.org/10.1016/S0370-2693(97)00850-2}{\emph{Phys. Lett. B}
  {\bfseries 409} (1997) 387}
  [\href{https://arxiv.org/abs/hep-lat/9607032}{{\ttfamily hep-lat/9607032}}].

\bibitem{Duncan:1996xy}
A.~Duncan, E.~Eichten and H.~Thacker, \emph{{Electromagnetic splittings and
  light quark masses in lattice QCD}},
  \href{https://doi.org/10.1103/PhysRevLett.76.3894}{\emph{Phys. Rev. Lett.}
  {\bfseries 76} (1996) 3894}
  [\href{https://arxiv.org/abs/hep-lat/9602005}{{\ttfamily hep-lat/9602005}}].

\bibitem{Hayakawa:2008an}
M.~Hayakawa and S.~Uno, \emph{{QED in finite volume and finite size scaling
  effect on electromagnetic properties of hadrons}},
  \href{https://doi.org/10.1143/PTP.120.413}{\emph{Prog. Theor. Phys.}
  {\bfseries 120} (2008) 413}
  [\href{https://arxiv.org/abs/0804.2044}{{\ttfamily 0804.2044}}].

\bibitem{Endres:2015gda}
M.G.~Endres, A.~Shindler, B.C.~Tiburzi and A.~Walker-Loud, \emph{{Massive
  photons: an infrared regularization scheme for lattice QCD+QED}},
  \href{https://doi.org/10.1103/PhysRevLett.117.072002}{\emph{Phys. Rev. Lett.}
  {\bfseries 117} (2016) 072002}
  [\href{https://arxiv.org/abs/1507.08916}{{\ttfamily 1507.08916}}].

\bibitem{Lucini:2015hfa}
B.~Lucini, A.~Patella, A.~Ramos and N.~Tantalo, \emph{{Charged hadrons in local
  finite-volume QED+QCD with C$^{*}$ boundary conditions}},
  \href{https://doi.org/10.1007/JHEP02(2016)076}{\emph{JHEP} {\bfseries 02}
  (2016) 076} [\href{https://arxiv.org/abs/1509.01636}{{\ttfamily
  1509.01636}}].

\bibitem{Davoudi:2018qpl}
Z.~Davoudi, J.~Harrison, A.~J\"uttner, A.~Portelli and M.J.~Savage,
  \emph{{Theoretical aspects of quantum electrodynamics in a finite volume with
  periodic boundary conditions}},
  \href{https://doi.org/10.1103/PhysRevD.99.034510}{\emph{Phys. Rev. D}
  {\bfseries 99} (2019) 034510}
  [\href{https://arxiv.org/abs/1810.05923}{{\ttfamily 1810.05923}}].

\bibitem{Bussone:2017xkb}
A.~Bussone, M.~Della~Morte and T.~Janowski, \emph{{Electromagnetic corrections
  to the hadronic vacuum polarization of the photon within QED$_{\rm L}$ and
  QED$_{\rm M}$}},
  \href{https://doi.org/10.1051/epjconf/201817506005}{\emph{EPJ Web Conf.}
  {\bfseries 175} (2018) 06005}
  [\href{https://arxiv.org/abs/1710.06024}{{\ttfamily 1710.06024}}].

\bibitem{Asmussen:2016lse}
N.~Asmussen, J.~Green, H.B.~Meyer and A.~Nyffeler, \emph{{Position-space
  approach to hadronic light-by-light scattering in the muon $g-2$ on the
  lattice}}, \href{https://doi.org/10.22323/1.256.0164}{\emph{PoS} {\bfseries
  LATTICE2016} (2016) 164} [\href{https://arxiv.org/abs/1609.08454}{{\ttfamily
  1609.08454}}].

\bibitem{Feng:2018qpx}
X.~Feng and L.~Jin, \emph{{QED self energies from lattice QCD without power-law
  finite-volume errors}},
  \href{https://doi.org/10.1103/PhysRevD.100.094509}{\emph{Phys. Rev. D}
  {\bfseries 100} (2019) 094509}
  [\href{https://arxiv.org/abs/1812.09817}{{\ttfamily 1812.09817}}].

\bibitem{Desiderio:2020oej}
A.~Desiderio et~al., \emph{{First lattice calculation of radiative leptonic
  decay rates of pseudoscalar mesons}},
  \href{https://doi.org/10.1103/PhysRevD.103.014502}{\emph{Phys. Rev. D}
  {\bfseries 103} (2021) 014502}
  [\href{https://arxiv.org/abs/2006.05358}{{\ttfamily 2006.05358}}].

\bibitem{Frezzotti:2020bfa}
R.~Frezzotti, M.~Garofalo, V.~Lubicz, G.~Martinelli, C.T.~Sachrajda,
  F.~Sanfilippo et~al., \emph{{Comparison of lattice QCD+QED predictions for
  radiative leptonic decays of light mesons with experimental data}},
  \href{https://doi.org/10.1103/PhysRevD.103.053005}{\emph{Phys. Rev. D}
  {\bfseries 103} (2021) 053005}
  [\href{https://arxiv.org/abs/2012.02120}{{\ttfamily 2012.02120}}].

\bibitem{Lubicz:2016xro}
V.~Lubicz, G.~Martinelli, C.T.~Sachrajda, F.~Sanfilippo, S.~Simula and
  N.~Tantalo, \emph{{Finite-Volume QED Corrections to Decay Amplitudes in
  Lattice QCD}}, \href{https://doi.org/10.1103/PhysRevD.95.034504}{\emph{Phys.
  Rev. D} {\bfseries 95} (2017) 034504}
  [\href{https://arxiv.org/abs/1611.08497}{{\ttfamily 1611.08497}}].

\bibitem{DiCarlo:2021apt}
M.~Di~Carlo, M.T.~Hansen, A.~Portelli and N.~Hermansson-Truedsson,
  \emph{{Relativistic, model-independent determination of electromagnetic
  finite-size effects beyond the pointlike approximation}},
  \href{https://doi.org/10.1103/PhysRevD.105.074509}{\emph{Phys. Rev. D}
  {\bfseries 105} (2022) 074509}
  [\href{https://arxiv.org/abs/2109.05002}{{\ttfamily 2109.05002}}].

\bibitem{Bijnens:2019ejw}
J.~Bijnens, J.~Harrison, N.~Hermansson-Truedsson, T.~Janowski, A.~Jüttner and
  A.~Portelli, \emph{{Electromagnetic finite-size effects to the hadronic
  vacuum polarization}},
  \href{https://doi.org/10.1103/PhysRevD.100.014508}{\emph{Phys. Rev. D}
  {\bfseries 100} (2019) 014508}
  [\href{https://arxiv.org/abs/https://arxiv.org/abs/1903.10591}{{\ttfamily
  https://arxiv.org/abs/1903.10591}}].

\bibitem{Hermansson-Truedsson:2024mey}
N.~Hermansson-Truedsson, \emph{{Structure-dependent electromagnetic
  finite-volume effects to the hadronic vacuum polarisation}},  in \emph{{41st
  International Symposium on Lattice Field Theory}}, 8, 2024
  [\href{https://arxiv.org/abs/2408.08042}{{\ttfamily 2408.08042}}].

\bibitem{Colangelo:2005gd}
G.~Colangelo, S.~Durr and C.~Haefeli, \emph{{Finite volume effects for meson
  masses and decay constants}},
  \href{https://doi.org/10.1016/j.nuclphysb.2005.05.015}{\emph{Nucl. Phys. B}
  {\bfseries 721} (2005) 136}
  [\href{https://arxiv.org/abs/hep-lat/0503014}{{\ttfamily hep-lat/0503014}}].

\bibitem{Colangelo:2006mp}
G.~Colangelo and C.~Haefeli, \emph{{Finite volume effects for the pion mass at
  two loops}},
  \href{https://doi.org/10.1016/j.nuclphysb.2006.03.010}{\emph{Nucl. Phys. B}
  {\bfseries 744} (2006) 14}
  [\href{https://arxiv.org/abs/hep-lat/0602017}{{\ttfamily hep-lat/0602017}}].

\bibitem{Bijnens:2014dea}
J.~Bijnens and T.~R\"ossler, \emph{{Finite Volume at Two-loops in Chiral
  Perturbation Theory}},
  \href{https://doi.org/10.1007/JHEP01(2015)034}{\emph{JHEP} {\bfseries 01}
  (2015) 034} [\href{https://arxiv.org/abs/1411.6384}{{\ttfamily 1411.6384}}].

\bibitem{Bijnens:2017esv}
J.~Bijnens and J.~Relefors, \emph{{Vector two-point functions in finite volume
  using partially quenched chiral perturbation theory at two loops}},
  \href{https://doi.org/10.1007/JHEP12(2017)114}{\emph{JHEP} {\bfseries 12}
  (2017) 114} [\href{https://arxiv.org/abs/1710.04479}{{\ttfamily
  1710.04479}}].

\bibitem{Davoudi:2014qua}
Z.~Davoudi and M.J.~Savage, \emph{{Finite-Volume Electromagnetic Corrections to
  the Masses of Mesons, Baryons and Nuclei}},
  \href{https://doi.org/10.1103/PhysRevD.90.054503}{\emph{Phys. Rev. D}
  {\bfseries 90} (2014) 054503}
  [\href{https://arxiv.org/abs/1402.6741}{{\ttfamily 1402.6741}}].

\bibitem{DiCarlo:2024lue}
M.~Di~Carlo, \emph{{Isospin-breaking corrections to weak decays: the current
  status and a new infrared improvement}},
  \href{https://doi.org/10.22323/1.453.0120}{\emph{PoS} {\bfseries LATTICE2023}
  (2024) 120} [\href{https://arxiv.org/abs/2401.07666}{{\ttfamily
  2401.07666}}].

\bibitem{Tantalo:2016vxk}
N.~Tantalo, V.~Lubicz, G.~Martinelli, C.T.~Sachrajda, F.~Sanfilippo and
  S.~Simula, \emph{{Electromagnetic corrections to leptonic decay rates of
  charged pseudoscalar mesons: finite-volume effects}},
  \href{https://arxiv.org/abs/1612.00199}{{\ttfamily 1612.00199}}.

\bibitem{Gagliardi:2022szw}
G.~Gagliardi, F.~Sanfilippo, S.~Simula, V.~Lubicz, F.~Mazzetti, G.~Martinelli
  et~al., \emph{{Virtual photon emission in leptonic decays of charged
  pseudoscalar mesons}},
  \href{https://doi.org/10.1103/PhysRevD.105.114507}{\emph{Phys. Rev. D}
  {\bfseries 105} (2022) 114507}
  [\href{https://arxiv.org/abs/2202.03833}{{\ttfamily 2202.03833}}].

\bibitem{Giusti:2023pot}
D.~Giusti, C.F.~Kane, C.~Lehner, S.~Meinel and A.~Soni, \emph{{Methods for
  high-precision determinations of radiative-leptonic decay form factors using
  lattice QCD}}, \href{https://doi.org/10.1103/PhysRevD.107.074507}{\emph{Phys.
  Rev. D} {\bfseries 107} (2023) 074507}
  [\href{https://arxiv.org/abs/2302.01298}{{\ttfamily 2302.01298}}].

\bibitem{Frezzotti:2023ygt}
R.~Frezzotti, G.~Gagliardi, V.~Lubicz, G.~Martinelli, F.~Mazzetti,
  C.T.~Sachrajda et~al., \emph{{Lattice calculation of the $D_{s}$ meson
  radiative form factors over the full kinematical range}},
  \href{https://arxiv.org/abs/2306.05904}{{\ttfamily 2306.05904}}.

\bibitem{Bijnens:1992en}
J.~Bijnens, G.~Ecker and J.~Gasser, \emph{{Radiative semileptonic kaon
  decays}}, \href{https://doi.org/10.1016/0550-3213(93)90259-R}{\emph{Nucl.
  Phys. B} {\bfseries 396} (1993) 81}
  [\href{https://arxiv.org/abs/hep-ph/9209261}{{\ttfamily hep-ph/9209261}}].

\bibitem{Geng:2003mt}
C.Q.~Geng, I.-L.~Ho and T.H.~Wu, \emph{{Axial vector form-factors for K(l2
  gamma) and pi(l2 gamma) at O(p**6) in chiral perturbation theory}},
  \href{https://doi.org/10.1016/j.nuclphysb.2003.12.039}{\emph{Nucl. Phys. B}
  {\bfseries 684} (2004) 281}
  [\href{https://arxiv.org/abs/hep-ph/0306165}{{\ttfamily hep-ph/0306165}}].

\bibitem{PhysRevLett.85.2256}
{\scshape E787 Collaboration} collaboration, \emph{Measurement of
  structure-dependent
  ${\mathit{k}}^{+}\phantom{\rule{0ex}{0ex}}\ensuremath{\rightarrow}\phantom{\rule{0ex}{0ex}}{\mathit{\ensuremath{\mu}}}^{+}{\ensuremath{\nu}}_{\ensuremath{\mu}}\mathit{\ensuremath{\gamma}}$
  decay}, \href{https://doi.org/10.1103/PhysRevLett.85.2256}{\emph{Phys. Rev.
  Lett.} {\bfseries 85} (2000) 2256}.

\bibitem{PhysRevLett.89.061803}
A.A.~Poblaguev, R.~Appel, G.S.~Atoyan, B.~Bassalleck, D.R.~Bergman, N.~Cheung
  et~al., \emph{Experimental study of the radiative decays
  ${K}^{+}\ensuremath{\rightarrow}{\ensuremath{\mu}}^{+}\ensuremath{\nu}{e}^{+}{e}^{\ensuremath{-}}$
  and
  ${K}^{+}\ensuremath{\rightarrow}{e}^{+}\ensuremath{\nu}{e}^{+}{e}^{\ensuremath{-}}$},
  \href{https://doi.org/10.1103/PhysRevLett.89.061803}{\emph{Phys. Rev. Lett.}
  {\bfseries 89} (2002) 061803}.

\bibitem{KLOE:2009urs}
{\scshape KLOE} collaboration, \emph{{Precise measurement of $\Gamma(K \to e
  \nu(\gamma)) / \Gamma(K \to \mu \nu(\gamma))$ and study of $K \to e \nu
  \gamma$}}, \href{https://doi.org/10.1140/epjc/s10052-009-1217-6}{\emph{Eur.
  Phys. J. C} {\bfseries 64} (2009) 627}
  [\href{https://arxiv.org/abs/0907.3594}{{\ttfamily 0907.3594}}].

\bibitem{Tchikilev:2010wy}
O.~Tchikilev et~al., \emph{{Observation of the destructive interference in the
  radiative kaon decay $K\rightarrow \mu \nu \gamma$}},
  \href{https://arxiv.org/abs/1001.0374}{{\ttfamily 1001.0374}}.

\bibitem{DUK201159}
V.~Duk, V.~Bolotov, V.~Lebedev, A.~Khudyakov, A.~Makarov, A.~Polyarush et~al.,
  \emph{{Extraction of kaon formfactors from $K\rightarrow \mu \nu \gamma $
  decay at ISTRA+ setup}},
  \href{https://doi.org/https://doi.org/10.1016/j.physletb.2010.10.043}{\emph{Physics
  Letters B} {\bfseries 695} (2011) 59}.

\bibitem{Cirigliano:2011ny}
V.~Cirigliano, G.~Ecker, H.~Neufeld, A.~Pich and J.~Portoles, \emph{{Kaon
  Decays in the Standard Model}},
  \href{https://doi.org/10.1103/RevModPhys.84.399}{\emph{Rev. Mod. Phys.}
  {\bfseries 84} (2012) 399} [\href{https://arxiv.org/abs/1107.6001}{{\ttfamily
  1107.6001}}].

\bibitem{QedFvCoef}
M.~Di~Carlo and A.~Portelli. \url{https://github.com/aportelli/QedFvCoef},
  2023.

\bibitem{Hermansson-Truedsson:2023krp}
N.~Hermansson-Truedsson, M.~Di~Carlo, M.T.~Hansen and A.~Portelli,
  \emph{{Structure-dependent electromagnetic finite-volume effects through
  order $1/L^3$}}, \href{https://doi.org/10.22323/1.453.0265}{\emph{PoS}
  {\bfseries LATTICE2023} (2024) 265}
  [\href{https://arxiv.org/abs/2310.13358}{{\ttfamily 2310.13358}}].

\bibitem{ParticleDataGroup:2024cfk}
{\scshape Particle Data Group} collaboration, \emph{{Review of particle
  physics}}, \href{https://doi.org/10.1103/PhysRevD.110.030001}{\emph{Phys.
  Rev. D} {\bfseries 110} (2024) 030001}.

\end{thebibliography}\endgroup

\end{document}